\documentclass[%
 reprint,
superscriptaddress,
nofootinbib,
 amsmath,
 amssymb,
 aps,
 twocolumn,
]{revtex4-2}

\usepackage{subfigure}
\usepackage{siunitx}
\usepackage{graphicx}
\usepackage{xcolor}
\usepackage{xspace}
\usepackage{hyperref}
\usepackage[capitalize]{cleveref}

\newcommand{\massratio}{\ensuremath{q}\xspace}
\newcommand{\mchirp}{\ensuremath{\mathcal{M}^{\rm det}}\xspace}

\newcommand{\primaryspin}{\ensuremath{a_1}\xspace}
\newcommand{\Msun}{\ensuremath{M_{\odot}}\xspace}

\newcommand{\CHECK}[1]{{\color{red} #1}}

\newcommand{\data}{\ensuremath{\mathbf{d}}}
\newcommand{\likelihood}{\ensuremath{\mathcal{L}}}
\newcommand{\prior}{\ensuremath{\pi}}
\newcommand{\evidence}{\ensuremath{\mathcal{Z}}}
\newcommand{\pastro}{\ensuremath{p_{\rm astro}}\xspace}

\newcommand{\bilbyMCMC}{\textsc{Bilby-MCMC}\xspace}
\newcommand{\dynesty}{\textsc{dynesty}\xspace}
\newcommand{\bilby}{\textsc{Bilby}\xspace}
\newcommand{\pycbc}{\textsc{PyCBC}\xspace}
\newcommand{\gstlal}{\textsc{GstLAL}\xspace}

\newcommand{\pasa}{Publications of the Astronomical Society of Australia}
\newcommand{\mnras}{Monthly Notices of the Royal Astronomical Society}
\newcommand{\apjs}{APJS}
\newcommand{\apjl}{APJL}

\newcommand{\pvtwo}{\texttt{IMRPhenomPv2}\xspace}

\newcommand{\ofour}{GW190403\_051519\xspace}
\newcommand{\onine}{GW190929\_012149\xspace}

\begin{document}

\title[]{Parameterised population models of transient non-Gaussian noise in the LIGO gravitational-wave detectors}

\author{Gregory Ashton}
\address{University of Portsmouth, Institute of Cosmology and Gravitation, Portsmouth PO1 3FX, United Kingdom}
\address{Department of Physics, Royal Holloway, University of London, TW20 0EX, United Kingdom}
 \email{gregory.ashton@ligo.org}
\author{Sarah Thiele}
\address{University of British Columbia, Vancouver, BC V6T 1Z4, Canada}
\author{Yannick Lecoeuche}
\address{LIGO Hanford Observatory, Richland, WA 99352, USA}
\author{Jess McIver}
\address{University of British Columbia, Vancouver, BC V6T 1Z4, Canada}
\author{Laura K Nuttall}
\address{University of Portsmouth, Institute of Cosmology and Gravitation, Portsmouth PO1 3FX, United Kingdom}

\begin{abstract}
The two interferometric LIGO gravitational-wave observatories provide the most sensitive data to date to study the gravitational-wave Universe.
As part of a global network, they have just completed their third observing run in which they observed many tens of signals from merging compact binary systems.
It has long been known that a limiting factor in identifying transient gravitational-wave signals is the presence of transient non-Gaussian noise, which reduce the ability of astrophysical searches to detect signals confidently.
Significant efforts are taken to identify and mitigate this noise at the source, but its presence persists, leading to the need for software solutions. Taking a set of transient noise artefacts categorised by the GravitySpy software during the O3a observing era, we produce parameterised population models of the noise projected into the space of astrophysical model parameters of merging binary systems.
We compare the inferred population properties of transient noise artefacts with observed astrophysical systems from the GWTC2.1 catalogue.
We find that while the population of astrophysical systems tend to have near equal masses and moderate spins, transient noise artefacts are typically characterised by extreme mass ratios and large spins.
This work provides a new method to calculate the consistency of an observed candidate with a given class of noise artefacts.
This approach could be used in assessing the consistency of candidates found by astrophysical searches (i.e. determining if they are consistent with a known glitch class).
Furthermore, the approach could be incorporated into astrophysical searches directly, potentially improving the reach of the detectors, though only a detailed study would verify this.

\end{abstract}

\maketitle

\section{Introduction}
The LIGO, Virgo, and KAGRA gravitational-wave detectors \citep{aasi_2015, acernese_2015, aso_2013} have opened the door to the Gravitational-Wave Universe and provided our first insights into the mergers of black holes and neutron stars. These kilometer-scale interferometers, sensitive to signals in the tens of Hz to kHz band, survey the sky nearly isotropically during \emph{observing runs} \citep{abbott_2020_obs_scen}. 
Between each observing run, the detectors are improved, reducing the level of noise and yielding improvements in the sensitivity.
Therefore, each new observing run probes deeper than before, yielding an increase in the rate of detections.

The LIGO detectors contain noise which is often assumed to be stationary and Gaussian for analysis, but often exhibits non-stationarity and frequent non-Gaussian transient noise artefacts termed \emph{glitches} \citep{blackburn_2008, abbott_2016_transient_noise, cabero_2019, abbott_2020_guide, davis_2021}.
All glitches are believed to be instrumental or environmental in origin.
In some cases, the cause of a glitch class can be identified and improvements made to the detector to reduce or eradicate the class.
In other cases, so-called witness channels can identify when such glitches occurred and veto coincident candidate events (see, e.g. \citet{davis_2021}).
Finally, glitches in classes without a known origin or witness channel remain a feature of the science-mode data from interferometric gravitational-wave observatories.

Glitches are detrimental to achieving the scientific goals of gravitational-wave observatories.
Suppose glitches occur in coincidence with an actual astrophysical signal. In that case, this may result in the signal being missed by searches, reducing the number of detectable signals, or biasing our astrophysical inferences about the signal itself.
However, glitches that do not occur in coincidence with real signals also reduce the sensitivity of the astrophysical searches.
To identify signals, \emph{search pipelines} use information about waveform morphology and coincidences between detectors (see, e.g. \citep{allen_2012, cannon_2013, nitz_2017, aubin_2021, sachdev_2019}) to calculate a detection statistic.
The presence of glitches in the detector data produces values of the detection statistic larger than the expectation from Gaussian noise alone.
To estimate the non-astrophysical background of the detection statistic due to glitches, most search pipelines use a bootstrap approach, for example, by time-shifting the data from independent detectors.
These approaches enable a significance to be attached to a candidate (for example, by the false alarm rate or FAR).
Therefore, a candidate's significance is affected both by how alike it is to predictions for astrophysical signals and the level of background noise created by glitches.
Vetoing glitches or utilising detection statistics that better identify transient noise outliers (see, e.g. \citep{allen_2005}) reduce the level of background noise from glitches and hence improve the number of sources that can be identified at a given confidence level.
But, in either case, the presence of glitches makes it more challenging to identify astrophysical signals.

Significant effort has been made to classify glitches into separate \emph{classes} based on their morphology \citep{zevin_2017, mukherjee_2010, rampone_2013, powell_2015, powell_2017, mukund_2017, cabero_2019}.
Glitch classification can potentially identify the cause of glitches enabling detector improvements to reduce their impact.
Classification can also help mitigate the impact of glitches on a search, for example, by building a glitch-robust detection statistic \citep{jadhav_2020} or help qualitatively interpret putative astrophysical signals by placing them in the context of known glitch classes.

\citet{davis_2020} investigated the impact of four glitch classes, Blips, Koi Fish, Scattered Light, and Scratchy glitches on the \pycbc search pipeline \citep{nitz_2017}.
Calculating a rate at which the glitches pass a given detection threshold over the physical parameter space of sources, they demonstrate a method to calculate the significance of a possible signal that coincides with a glitch.
An alternative approach explored in \citet{ashton_2019b} builds information about the distribution of glitches into a Bayesian odds, eschewing a bootstrapped estimation of the background.
However it is done, detection approaches which reduce the impact of glitches are critical to
improving the search sensitivity and hence identifying more astrophysical candidates.

In this work, we take a modelled approach to glitch classification.
We aim to deliver simple probabilistic models for the glitch classes most harmful to astrophysical analyses.
To do this, we analyse single-detector glitch triggers pre-classified by the GravitySpy citizen science project \citep{zevin_2017} from the LIGO Hanford and Livingston detectors \citep{aasi_2015}.
We analyse each glitch using a model of a precessing binary black hole merger (see \citet{merritt_2021} for an alternative approach developing parametric glitch models).
Using an astrophysical black hole merger model enables us to project the properties of the glitch into the physical parameter space of astrophysical signals.
Taking these projections, we then build simple parameterised population models for each glitch class and study this population with respect to known astrophysical signals.

The rest of this paper is structured as follows.
In Sec.~\ref{sec:methodology}, we introduce the methodology based on the principles of population inference.
In Sec.~\ref{sec:results}, we describe our results fitting population models to 4 different glitch classes.
In Sec.~\ref{sec:comparison}, we compare the glitch population with the population of binary black-hole mergers described in the GWTC2.1 transient gravitational-wave catalogue.
Finally, we conclude in Sec.~\ref{sec:conclusion}.

\section{Methodology}
\label{sec:methodology}

\begin{figure}
    \centering
    \includegraphics[width=0.5\textwidth]{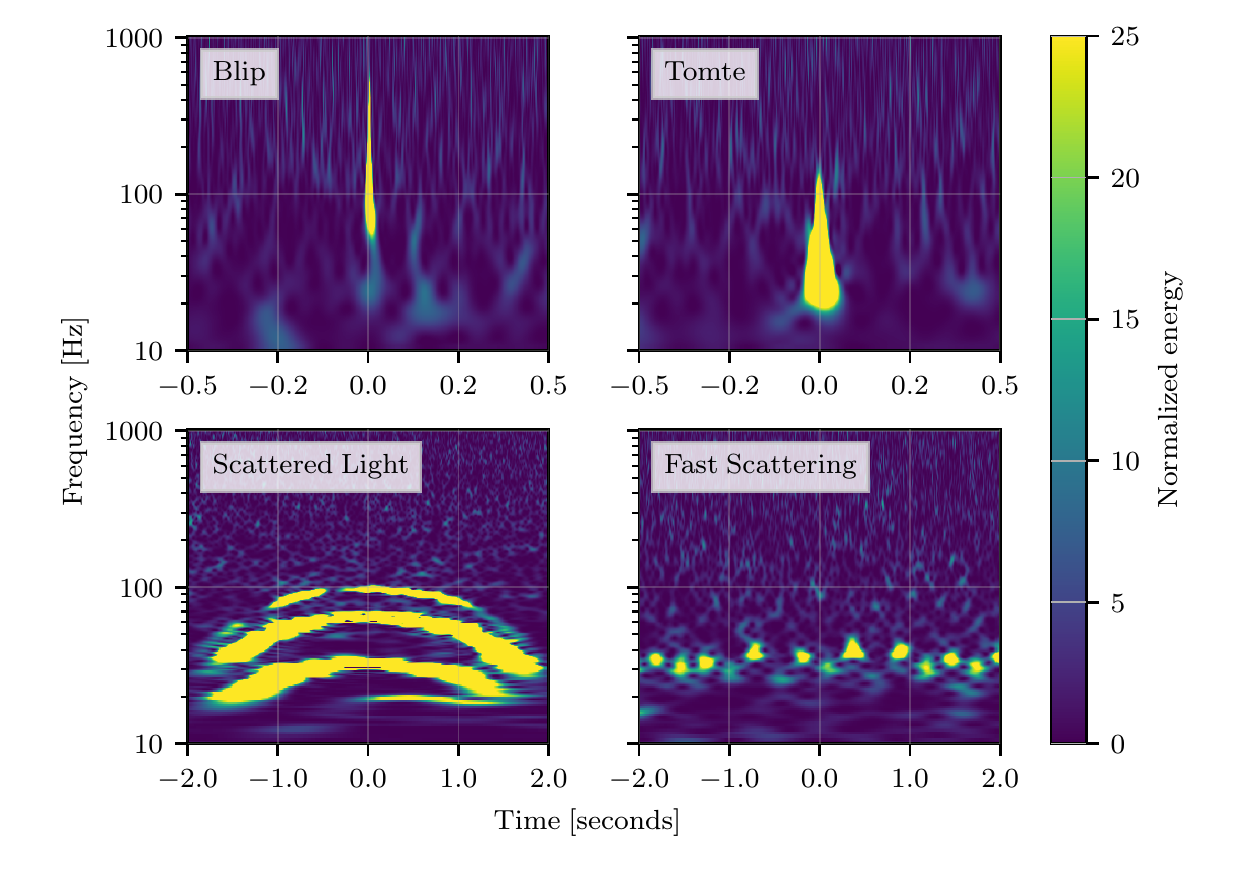}
    \caption{Time-frequency spectrograms of the four glitch classes which we analyse in this work.}
    \label{fig:glitch_examples}
\end{figure}

The methodology used herein follows the principles of \emph{hierarchical population inference} for compact binary coalescence (CBC) signals (see Refs.~\citep{mandel_2010, mandel_ros_2010, adams_2012} and \citet{thrane_2019} for a review). We aim to identify the population properties of 4 glitch classes \emph{Blip}, \emph{Tomte}, \emph{Fast Scattering}, and \emph{Scattered Light} which are known to most adversely impact search sensitivity \citep{davis_2020, davis_2021}.
In \cref{fig:glitch_examples}, we provides time-frequency spectrograms demonstrating typical examples of each glitch class.
These classes were the most numerous during the O1 to O3 observing runs and will likely persist for future observing runs.
We perform the population inference separately on glitches in the LIGO Hanford (H1) and LIGO Livingston (L1) detectors. This produces two sets of population properties that we can use to understand the consistency of glitch classes between the detectors.

To infer the properties of glitches in a given class, we first identify a list of \emph{glitch times} pre-categorised by GravitySpy with a confidence threshold of 95\%.
We check that these glitches would typically contribute to the noise backgrounds estimated by search pipelines, i.e. they would not be vetoed by the single-detector signal-to-noise ratio (SNR) cuts; the details of this are given in the subsections of Sec.~\ref{sec:results}.
Each glitch is individually vetted by eye using a time-frequency visualisation to ensure it is consistent with the corresponding glitch class.
We then analyse the data $\data_i$ surrounding the $i^{\rm th}$ glitch in our list using a stochastic sampling algorithm (specifically, the nested sampling \dynesty~\citep{speagle_2020} algorithm through the \bilby~\citep{ashton_2019} package) and a model $M$.
In this work, we use the \pvtwo \citep{hannam_2014, schmidt_2012} model which includes the inspiral, merger and ringdown of merging binary black holes and has been a standard benchmark waveform for several observing runs \citep{GWTC-1_2019, GWTC2_2021}. 
\pvtwo, however, only models the dominant $(\ell, m)=(2, 2)$ modes in the gravitational-wave radiation.
Recent advances in waveform modelling have produced improved models, including the effect of higher-order modes \citep{london_2018, cotesta_2018, nagar_2018, varma_2019, pratten_2021, estelles_2020}.
Nevertheless, we use \pvtwo as it is well standardised and fast to evaluate, enabling us to analyse hundreds of glitches without a high computational cost.
However, this implies that we infer the properties of glitches given the \pvtwo waveform model.
It is well known that systematic uncertainty exists between waveform models and that models including higher-order modes produce quite different posterior distributions, especially when the system has asymmetric masses \citep{GW190412_discovery, GW190521_discovery, GW190814_discovery}.
We, therefore, caution when comparing the glitch population obtained herein with candidate triggers that are analysed using waveforms other than \pvtwo, especially if those waveforms include the effect of higher-order modes.

The \pvtwo model has an associated set of model parameters $\theta$, which describe the mass, spins, location, and orientation of the source.
For each glitch class, we produce an estimate of the posterior distribution $p(\theta | \data_i, M)$ (in the form of a set of posterior samples) and the Bayesian evidence $\evidence(\data_i | M)$.
During the analysis of individual glitches, we use a Bayesian prior $\prior(\theta)$ appropriate for analysing astrophysical binary black holes.
Specifically, the prior is uniform in the detector-frame chirp mass\footnote{We infer the glitch population properties of the chirp mass as measured in the detector frame, \mchirp, rather than the source-frame $\mathcal{M}=\mchirp / (1 + z)$ where $z$ is the inferred redshift} \mchirp \citep{cutler_1994} between 8 and 200 \Msun and uniform in mass ratio $q$ between $1/20$ and 1 and assumes an isotropic distribution on the black holes spins. For all other parameters, we use the standard non-informative astrophysical prior distributions defined in \citet{veitch_2015}.

To infer the population properties of glitches in a given class, we take the set of posterior samples and Bayesian evidence from each individual analysis and perform a hierarchical population inference.
Specifically, for a parameter $\theta_i \in \theta$, we define a population hyper-model $H$ with associated model hyperparameters $\Lambda$ such that we can evaluate $\pi(\theta_i| \Lambda, H)$.
Once defined, we then utilise a variation on importance sampling to construct the likelihood $\likelihood(\{\data _i | \Lambda, H)$.
(We describe the population hypermodels and specifics of the likelihood used in this work below).
Given this likelihood and a suitable hyperprior $\pi(\Lambda | H)$, we then use stochastic sampling  to infer $p(\Lambda | \{\data\}, H)$, the population hyperparameters conditioned on all glitches in the given class.
For this analysis, we use the \bilbyMCMC~\citep{ashton_2021} Markov-Chain Monte-Carlo sampling algorithm; we verify that both the \dynesty and \bilbyMCMC samplers produce equivalent results, but that \bilbyMCMC is more efficient in this instance where only the posterior distribution and not the Bayesian evidence is of interest.

In principle, a hyper-model can predict the population behaviour for the full 15-dimensional parameter space $\theta$, including correlations between parameters.
However, for a sub-set of parameters, we do not observe any population trends (e.g., the sky localisation of a population of glitch events tends to be isotropically distributed).
We find that three parameters, the chirp mass $\mchirp$, mass ratio $\massratio$, and the dimensionless spin-magnitude of the primary (heavier-mass) object $\primaryspin$ are the primary indicators for the glitch classes considered in this work. We will build population models for each of these parameters independently.
There is a weak correlation between these parameters, but we choose to infer their properties separately.
This discards information about the correlation but greatly simplifies the population models.
In future work, we will investigate whether including such correlations can produce a more accurate probabilistic model.

For other parameters, we do not observe strong trends in the population, except for the detector-frame luminosity distance, for which we observe a piling up of events at the lower bound (100~Mpc).
The glitches selected for this analysis are, by design, loud glitches which pass a confidence threshold for categorisation by the GravitySpy (see \citet{zevin_2017} for details).
Hence, our set of glitches has an intrinsic selection bias in the loudness and hence luminosity distance.
It is no surprise then that we observe a trend that all glitches tend to have small luminosity distances, as this is the dominant term in the signal amplitude and hence determines how ``loud'' a glitch is.
We choose not to model the luminosity distance as we would need to properly calibrate the model for the intrinsic selection bias of our set of glitches.
Other parameters, such as the system mass, may also be affected by the selection bias, but we model them nonetheless.

In this work, we will use three hypermodels to capture the broad-scale population properties of each glitch class for each parameter.
These are three hypermodels from an infinite set of possible models.
We arrived at these by studying the performance of a larger superset of hypermodels using posterior-predictive checks (described below).
Where more complicated models failed to improve the posterior-predictive checks, we used simpler models. 
These hyper-models are designed to capture the broad features in a simple parameterised form.
The three hypermodels are:
\begin{enumerate}
    \item [M1:] A \emph{power law} distribution, it has a single hyperparameter $\Lambda=\{\alpha\}$ and density
    \begin{align}
        \pi(\theta_i | \alpha, H_1) \propto \theta_i^{\alpha}\,,
    \end{align}
    with bounds given by the physical bounds on $\theta_i$. This distribution is useful for modelling populations which tend to \emph{rail} against the physical bounds (for example, the primary spin). For $\alpha$, we use a normally distributed prior with zero mean and a standard deviation of 5. This removes the requirement for an arbitrary upper bound on the magnitude while exponentially suppressing very large values. 
    \item [M2:] A truncated \emph{skew-normal} distribution \citep{ohagan_1976, azzalini_1999}, it has three hyperparameters $\Lambda=\{\mu, \sigma, \kappa\}$ and density
    \begin{align}
        \pi(\theta_i | \mu, \sigma, \kappa, H_2) = \frac{2}{\sigma} \phi\left(\frac{\theta_i - \mu}{\sigma}\right)\Phi\left(\kappa \frac{\theta_i - \mu}{\sigma}\right)
    \end{align}
    where $\phi(x)$ is the standard normal probability density function and $\Phi(x)$ is the error function. We truncate the distribution again to the physical bounds on $\theta_i$. This distribution is useful for modelling unimodal populations with a distinct peak.
    We use a uniform prior for $\mu$ over the prior support on $\theta_i$, a normal prior for $\kappa$ with zero mean and a standard deviation of 10 and a truncated normal prior for $\sigma$ with standard deviation given by the maximum support on $\theta_i$.
    \item [M3:] A mixture-model of two normal distributions with five hyperparameters $\Lambda=\{\xi, \mu_0, \sigma_0, \mu_1, \sigma_1\}$ and density
    \begin{align}
        \pi(\theta_i | \Lambda)
        = \frac{\xi}{\sigma_0} \phi\left(\frac{\theta_i - \mu_0}{\sigma_0}\right)
        + \frac{1 - \xi}{\sigma_1} \phi\left(\frac{\theta_i - \mu_1}{\sigma_1}\right)\,.
    \end{align}
    This is effective in modelling multi-modal distributions; we do not include a skew as this was found to be poorly constrained in the cases where the M3 model was required.
    As is done for the M2 model, we use a uniform prior for $\mu_0$ and $\mu_1$ and a truncated normal distribution for $\sigma_0$ and $\sigma_1$.
    For $\xi$, we apply a uniform prior on [0, 0.5].
\end{enumerate}
For each model, the likelihood is constructed from Equation (32) of \citet{thrane_2019} using ``recycling''.
This enables a two-step approach (first analysing individual events, then the population properties), which is critical to reducing the wall-time of the analysis.
For each hyper-model, we report the median values of $p(\Lambda | \{\data_i\}, H)$. These, together with the model descriptions above, constitute the core product of this work: simple parameterised models of glitches in the LIGO detectors.

To verify that the hyper-model captures the broad-scale features of the population, we use predictive posterior checking.
First, to plot the ``Measured'' population, we bin each individual glitch-posterior in a histogram, then plot the mean and the $90\%$ credible interval (between 5\% and 95\%).
This provides a summary of the population properties, which includes the variation in the individual posteriors\footnote{Some alternatives, e.g. a histogram of the means, exclude information about the posterior uncertainty which is captured by our method}.
An example of this population visualisation can be seen in the right-hand column of \cref{tab:blip}: the ``Measured'' curve shows the mean (solid grey line) and 90\% credible interval (filled grey region) for the population of Blip glitches across separate parameters.
Second, to plot the ``Predicted'' population, we repeat the process above but simulate the individual posteriors from the population model.
Specifically, we draw $\theta_i$ from $p(\theta_i | \Lambda')$ (where $\Lambda'$ is the median)\footnote{An improvement to the predictive check would be to draw $\Lambda'$ from the inferred posterior. However, we wish to understand the ability of our simple models, parameterised by the median of the hyperparameter distribution, to explain the data.} then simulate a posterior and repeat the steps above.
The simulated posterior is derived by randomly sampling a posterior from the ``Measured'' population and aligning its mean with a random draw from the posterior population distribution.
This is a non-optimal method, but the alternative (simulating the signal predicted by the posterior population distribution in real detector noise unaffected by other transient noise and performing Bayesian inference) is too computationally demanding to be feasible.
Examples of the resulting posterior predictive checks can be found in the right-hand column of \cref{tab:blip}: the ``Predicted'' curve shows the mean (solid orange line) and 90\% credible (filled orange region) predicted by the hyper model across separate parameters.
Here we see that the models capture the broad features (e.g. the number and location of modes and their width) but do not always capture the details (e.g. the Measured and Predicted curves do not overlap perfectly).

\section{Results}
\label{sec:results}
We now describe the results for each glitch class in the sections below.
For each glitch class, we provide a table with a summary (the glitch model and median inferred population parameters) and the results of the posterior predictive checks.
We also provide this information in a machine-readable table A \citep{ashton_gregory_2021_5550510}.
For each model, we report the results of a single hypermodel.
However, we fitted various models during development and used the posterior predictive checks as a qualitative guide to select the best fitting model.

\subsection{Tomte glitches}
\label{sec:tomte}
In \cref{tab:tomte}, we report the hypermodel, median, and posterior predictive checks for the population properties in \mchirp, \massratio, and \primaryspin for Tomte glitches in the H1 and L1 detectors.
For each detector, we analyse a set of 1000 Tomte glitches identified with GravitySpy.\footnote{Of these 1000, computational issues resulted in the failure to analyse 1 Tomte glitch from the L1 set; as such, our results pertain to 1000 H1 Tomte glitches and 999 L1 Tomte glitches}.
The SNR of these glitches ranges from 7.7 to 74; the distribution is peaked with a median SNR of 21, but 90\% of the Tomte glitches have an SNR less than 36.\footnote{Throughout, we refer to the glitch SNR as the value measured by the \texttt{Omicron}~\citep{robinet_2020} software. The lower bound of 7.5 is the result of a threshold applied when selecting the glitches for analysis.
}
We also verify that we find similar conclusions about the properties when we downsample to 100 glitches; we use this, later on, to justify analysing smaller sets of the Fast Scattering and Scattered Light glitches.

For the chirp mass, the measured glitch population is unimodal with a peak at $\sim\SI{75}{\Msun}$ and some positive skew.
We fit the population using the M2 Skew-Normal distribution and find a qualitatively acceptable fit for both detector populations. However, the model under-predicts the rate of glitches with $\mchirp\sim [125, 175]$ in the H1 detector.
While the broad features are consistent between H1 and L1, the amount of skew is noticeably different, and Tomte glitches in the H1 detectors have chirp mass values up to $\sim 150$, while Tomte glitches in the L1 detector, $\mchirp \lesssim 125$.

For the mass ratio, the population is unimodal with some skew and the bulk of support on $\massratio \lesssim 1/2$.
We fit the mass ratio distribution using the M2 Skew Normal model and find a good qualitative fit in H1 and L1.
As with the chirp mass, we note some qualitative differences between the H1 and L1 detectors with greater support in the H1 detectors for mass ratio's up to 0.5, which is not seen in the L1 glitch population.

The primary spin, \primaryspin piles up at the extreme spin case $\primaryspin=1$; we fit this using the M1 power-law model and find a reasonable agreement.
However, we note a lack of events at the $\primaryspin=1$ bound.
Our fitted model does not capture this feature; however, it is sufficiently narrow that overall the model remains a reasonable fit.
The power-law index $\alpha\sim 5-6$ provides a rough guide to the extent to which the spin ``piles up''; this value is less extreme than in the Blip glitch population discussed shortly.
For the spin, we do not observe significant differences between the H1 and L1 detector populations.

\subsection{Blip glitches}
In \cref{tab:blip}, we report the hypermodel, median, and posterior predictive checks for the population properties in \mchirp, \massratio, and \primaryspin for Blip glitches in the H1 and L1 detectors.
We analyse a set of 1000 Blip glitches in both H1 and L1 identified with GravitySpy.
\footnote{Of these 1000, computational issues resulted in the failure to analyse 61 Blip glitches from the L1 set; as such, our results pertain to 1000 H1 Blip glitches and 939 L1 Tomte glitches}.
The SNR of these glitches ranges from 7.5 to 47. The distribution peaks at the minimum bound, and 90\% of the glitches have an SNR less than 20.
We also verify that our results are robust when using only a subset of 100 randomly selected glitches.

The measured glitch population is unimodal for the chirp mass, with a peak at $\sim 25 \Msun$ and some positive skew.
We find that the M1 Skew-Normal model provides a good qualitative fit.
Comparing the measured and predicted posterior checks, the distribution means are well aligned, though the Skew-Normal under-predicts the density in the $40-60~\Msun$ region.

For the mass ratio, the population is bimodal, with identical features found in both detectors.
The bimodality originates from bimodal features observed in the posterior distributions of individual events. Though we see significant scatter in the shape of individual posteriors, averaging over multiple glitches, we observe similar distributions in both the H1 and L1 detectors indicating a common origin.
The bulk of the support for both modes is found in the region $\massratio \lesssim 1/2$, with some support up to the equal-mass bound.
We apply the M3 normal mixture model, which provides a reasonable fit to the data, though it tends to underpredict the number of nearly equal-mass glitches in the L1 detector.

Finally, for \primaryspin, the magnitude of the primary spin, we find that the measured glitch values pile up onto the extreme spin case $\primaryspin \sim 1$.
We also found this feature in the Tomte population, but here it is more extreme.
We find the simplistic M1 power-law model provides a suitable fit to this data; the extreme power-law index $\alpha \sim 37$ demonstrates the strength of the pile-up.

Across all three parameters, we find remarkably consistent glitch behaviour between the H1 and L1 detectors.
This, of course, may be simply due to the robust classification of events by GravitySpy.
However, we note that the Tomte glitches did not show such consistency.

\subsection{Fast Scattering glitches\label{sec:fast_scattering}}
In \cref{tab:fast_scattering}, we report the hypermodel, median, and posterior predictive checks for the population properties in \mchirp, \massratio, and \primaryspin for Fast Scattering glitches in the H1 and L1 detectors.
We use a set of 100 Fast Scattering glitches identified with GravitySpy.
We limit ourselves to 100 glitches based on our analysis of the Blip and Tomte glitches, which indicated 100 glitches is sufficient to represent the population properties.
This ten-fold reduction in events correspondingly reduces the computational burden of the analysis.
The SNR distribution of these glitches is strongly peaked at the minimum bound of 7.5, with 90\% of the glitches have an SNR less than 13. However, the distribution has a long tail with a maximum value of 121.

For the chirp mass, we find that the population has support from \SI{75}{\Msun} up to the artificial prior bound of \SI{200}{\Msun}.
This indicates that the actual population likely have support at larger masses still.
Such extreme heavy systems are beyond the space of likely astrophysical signals that ground-based detectors will observe in the advanced era.
As such, extending our prior beyond this artificial bound is unlikely to help distinguish glitches from signals in the near term, so we choose not to do so.
The observed distribution is ``lumpy'' with multiple components.
We choose to fit the distribution with a Skew Normal distribution.
This does not capture the ``lumpy'' behaviour but does broadly capture the typical support.

For the mass ratio, the Fast Scattering population have extreme mass ratios with a median of $\sim 1/10$ with some support up to $q\sim0.8$.
We fit this distribution using the M2 Skew Normal model and find a reasonable agreement between the predicted and measured distribution.

The primary spin, as with previously studied glitches, peaks at the extremal $\primaryspin=1$ bound.
However, the measured population has significant support for systems with $\primaryspin$ right down to zero.
We fit this distribution using the M1 power-law model and find a good fit.

\subsection{Scattered light glitches}
\label{sec:scatter_light}
In \cref{tab:scattered_light}, we report the hypermodel, median, and posterior predictive checks for the population properties in \mchirp, \massratio, and \primaryspin for Scattered Light glitches in the H1 and L1 detectors.
We use a set of 100 Scattered Light glitches identified with GravitySpy. The choice to use just 100 glitches mirrors the motivation in Sec.~\ref{sec:fast_scattering}.
The SNR of these glitches ranges from 7.5 to 370; as with the Fast Scattering glitch type, the distribution peaks at the lower bound with a long tail. 90\% of the glitches have an SNR less than 47.

For the chirp mass, we find that the population has support from \SI{75}{\Msun} up to the artificial prior bound of \SI{200}{\Msun} with significant ``lumpiness''.
The behaviour in the chirp mass distribution is similar in form between the Fast Scattering and Scattered Light glitches.
For the L1 detector, there is a significant overabundance at the equal-mass bound.
We choose to fit the distribution with a Skew Normal distribution.
This does not capture the ``lumpy'' behaviour but does broadly capture the typical support.

For the mass ratio, the Scatter Light population have extreme mass ratios with a median of $\sim 1/10$. But, unlike the Scattered Light glitches, $\massratio \lesssim 0.2$.
We fit this distribution using the M2 Skew Normal model and find a reasonable agreement between the predicted and measured distribution.

As with previously studied glitches, the primary spin peaks at the extremal $\primaryspin=1$ bound but with little support for smaller spins.
We fit this distribution using the M1 power-law model and find a good fit.
The power-law index $\alpha \sim 28$ is nearly as large as that of the Blip glitches.

\newcommand{\tfwidth}{0.29\textwidth}
\newcommand{\ici}[1]{\raisebox{-.5\height}{\includegraphics[trim=0mm 5mm 0mm -2mm, width=\tfwidth]{#1}}}

\section{Comparison with the GWTC2.1 catalogue}
\label{sec:comparison}
So far, \CHECK{57} gravitational-wave signals have been reported and analysed by the LIGO, Virgo and KAGRA collaborations \citep{GWTC-1_2019, GWTC2_2021, GWTC-2-1_2021, NSBH_detection}.
Predominantly, these have been binary black hole systems (BBH) except for two binary neutron star (BNS) and two neutron star black hole (NSBH) binaries.
Here we compare our glitch population with the \CHECK{47} events with $\pastro > 0.5$ reported in either the GWTC2 \citep{GWTC2_2021} or GWTC2.1 \citep{GWTC-2-1_2021} catalogues during the O3a observing run.\footnote{The GWTC2.1 catalogue reported on 44 events with $\pastro>0.5$ observed during O3a, we then added the 3 events reported in GWTC2, but which had a re-calculated $\pastro < 0.5$. This superset, therefore, includes all events during O3a reported by the collaboration with $\pastro > 0.5$.}

\begin{figure}[t]
    \centering
    \includegraphics[width=0.5\textwidth]{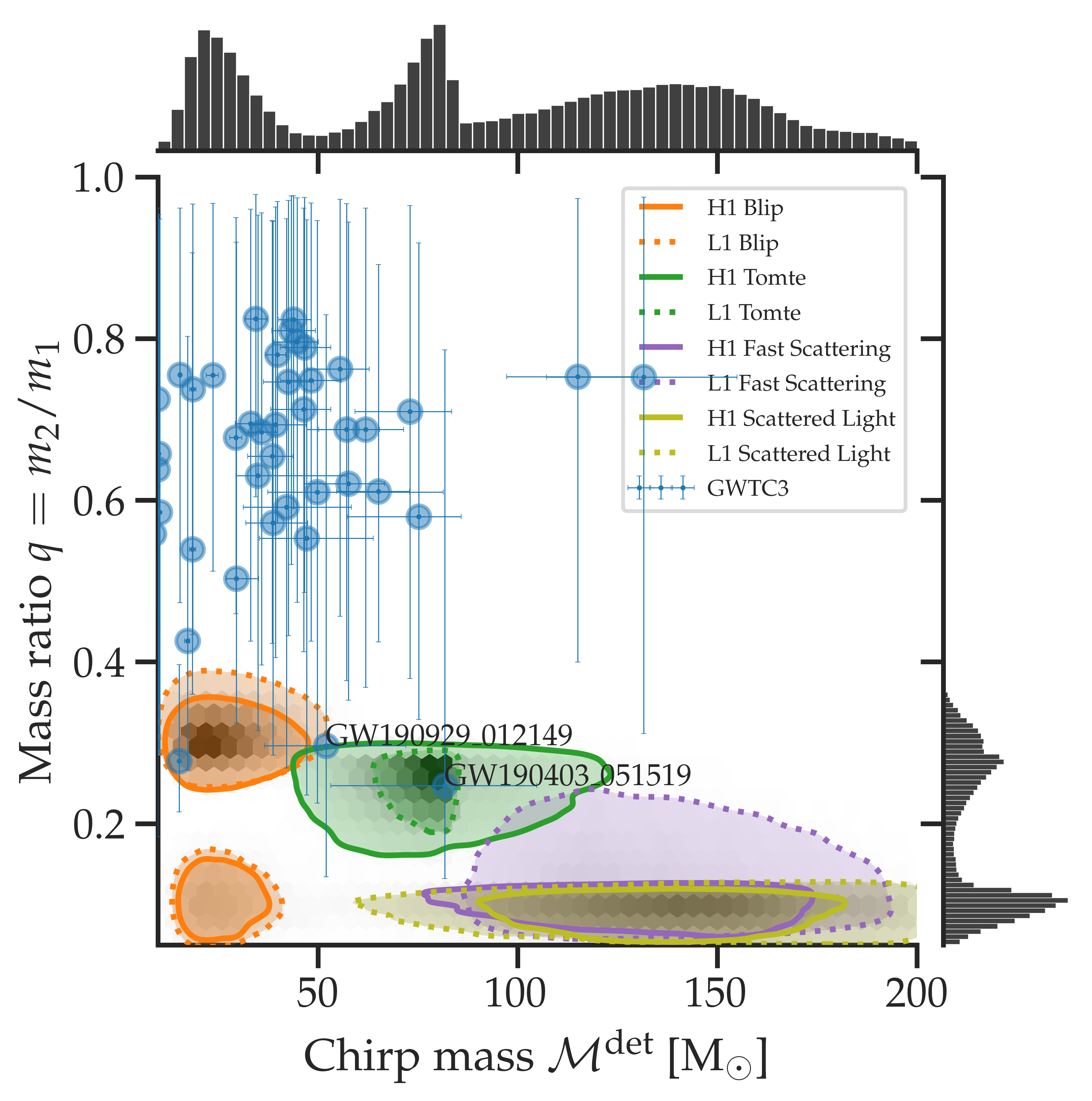}
    \caption{Visualisations of the parameterised glitch models developed in this work with the set of all events observed in the GWTC2 and GWTC2.1 catalogues \citep{GWTC2_2021, GWTC-2-1_2021} for the chirp mass and mass ratio.
    A solid (dashed) line marks the 90\% credible interval for the predicted distribution in the H1 (L1) detector, coloured by the glitch classification.
    A black density plot (and 1D marginal histograms) indicate the distribution of all glitches.
    Blue markers indicate all confident detections from the GWTC2 and GWTC2.1 catalogues.
    We remind the reader that the chirp mass reported here is measured in the red-shifted detector frame, not the source frame.
    }
    \label{fig:gwtc2_Mc_q}
\end{figure}

\begin{figure}[t]
    \centering
    \includegraphics[width=0.5\textwidth]{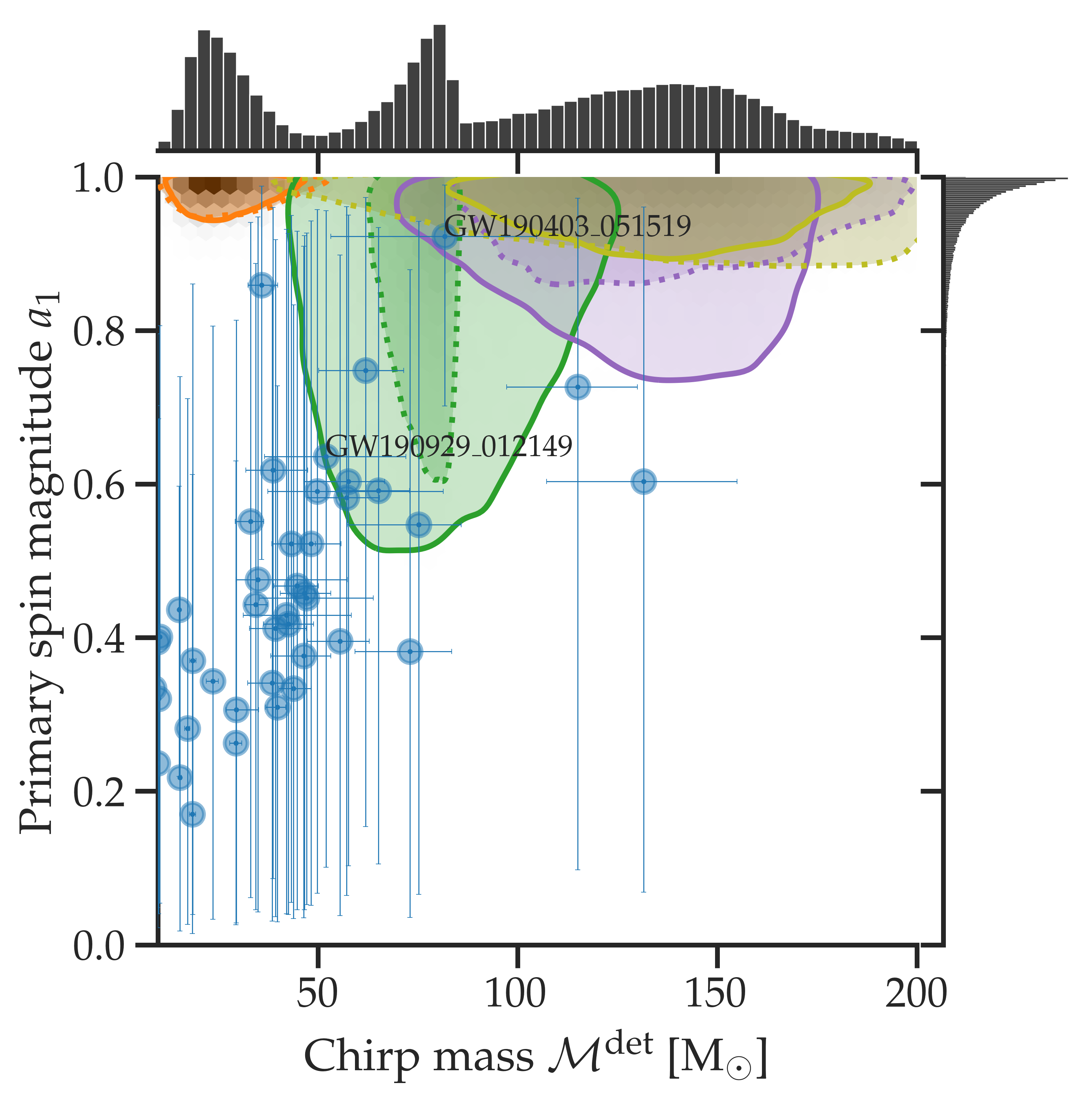}
    \caption{Visualisations of the parameterised glitch models developed in this work for the chirp mass and primary spin magnitude. See \cref{fig:gwtc2_Mc_q} for a complete description.}
    \label{fig:gwtc2_Mc_a1}
\end{figure}

\begin{figure}[t]
    \centering
    \includegraphics[width=0.5\textwidth]{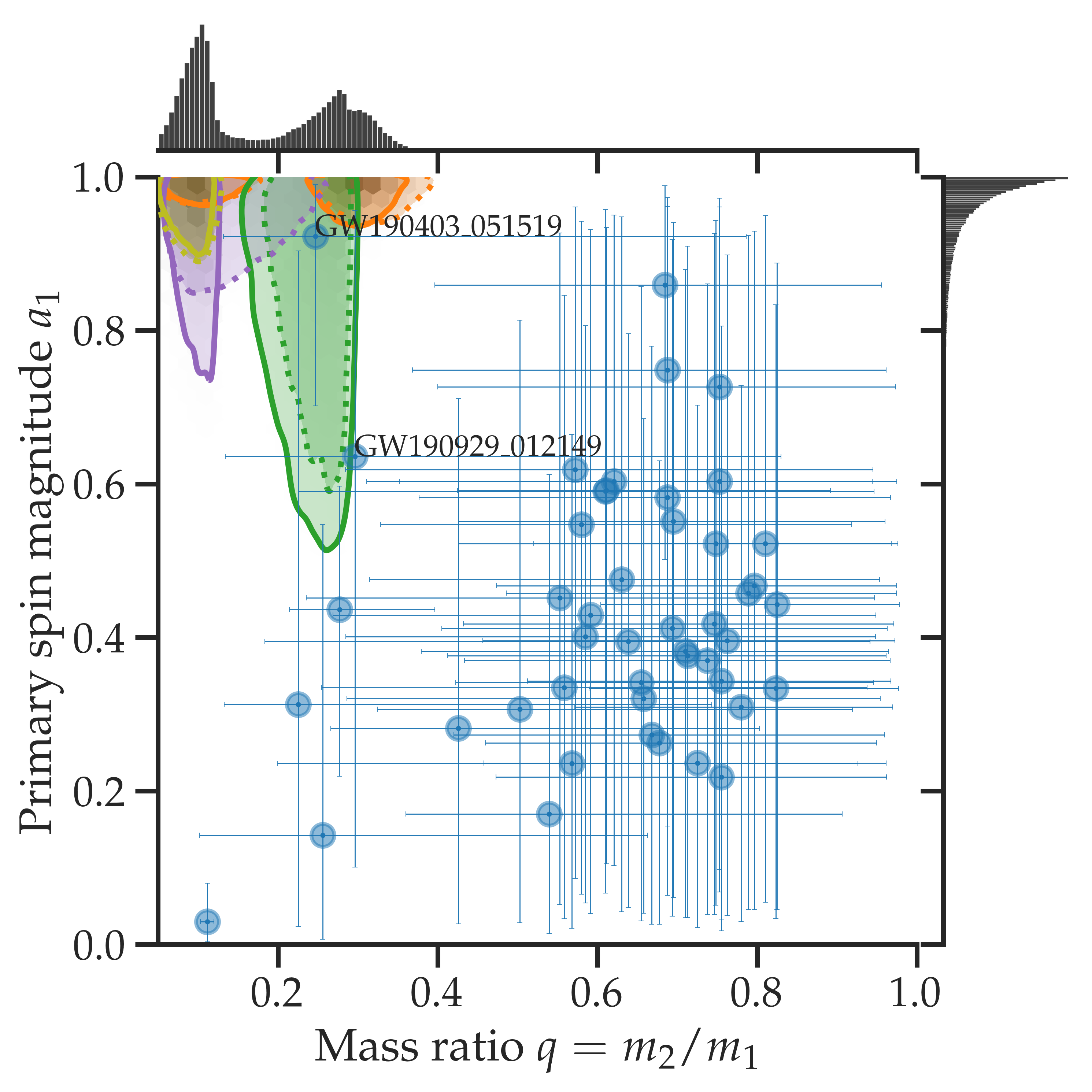}
    \caption{Visualisations of the parameterised glitch models developed in this work for the mass ratio and primary spin magnitude. See \cref{fig:gwtc2_Mc_q} for a complete description.}
    \label{fig:gwtc2_q_a1}
\end{figure}

We provide two-dimensional plots comparing our glitch population with the observed signal in \mchirp-\massratio (\cref{fig:gwtc2_Mc_q}), \mchirp-\primaryspin (\cref{fig:gwtc2_Mc_a1}), and \massratio-\primaryspin (\cref{fig:gwtc2_q_a1}).
In each plot, we give the 90\% credible interval of the four glitch classes (for both H1 and L1) along with the combined equal-weighted glitch population in black.
Projections of the overall glitch population are given for each one-dimensional plane.
Overlaid on this, we then plot the symmetric 90\% credible interval for all events with $\pastro>0.5$ reported in either the GWTC2 or GWTC2.1 catalogues.

Comparing the glitch population with the astrophysical BBH distribution, three features stand out: $i)$ glitches have large spin magnitudes while astrophysical signals have median spin magnitudes $\lesssim 0.9$; $ii)$ glitches have extreme mass ratios while astrophysical signals tend to have more equal-mass; and finally, $iii)$, the combined glitch population spans the entire chirp mass range considered in this work, though individual glitch classes tend to cluster.

Glitches have extreme mass ratios and spin because the CBC signals models are a poor fit to the glitch morphology (see, e.g. \citet{merritt_2021}). In the equal-mass zero-spin limit, the CBC waveform is sinusoidal with a characteristic monotonically increasing frequency evolution.
Non-equal mass ratios and spins modify this behaviour introducing ``twisting-up'' effects which produce irregular waveform morphology's.
Glitches are caused by terrestrial disturbances, which are not expected to characteristically chirp up in frequency. So, when fitting the CBC signal model, it is unsurprising that the best fits happen in the non-equal mass and large spin limits.

Of the BBH systems reported so far in O3a, just one, \ofour, falls firmly inside the 90\% credible interval for the Tomte population in chirp mass, mass ratio, and primary spin magnitude.
Additionally, \onine sits at the edge of the Tomte distribution, but no signals are consistent with the other 3 glitch classes.
Several other signals fall within the interval for the chirp mass and primary spin magnitude but are clearly distinguished when looking at the mass ratio and primary spin magnitude.\footnote{We note that the GWTC2.1 analysis of these events used the higher-order mode waveforms \texttt{IMRPhenomXPHM} \citep{pratten_2021} and \texttt{SEOBNRv4PHM} \citep{cotesta_2018}.
As previously discussed, adding higher-order mode content not modelled by the \pvtwo waveform used in this work can result in significant shifts in the posterior distribution.
However, we re-analysed \ofour using the \pvtwo waveform and found the event remained consistent with the Tomte population.
Between the GWTC2.1 analysis and our re-analysis, the median detector-frame chirp mass shifted from $86$ to $56$ \Msun, the mass ratio from $0.31$ to $0.19$, and the primary spin magnitude from $0.93$ to $0.85$.}
However, it should be noted that both \ofour and \onine have lower signal to noise ratios than any of the glitches included in our population analysis since we select only high-confidence glitches.
However, \citet{merritt_2021} argue that both quiet and loud glitches (of the same class) exhibit similar morphology.

\begin{figure}[t]
    \centering
    \includegraphics{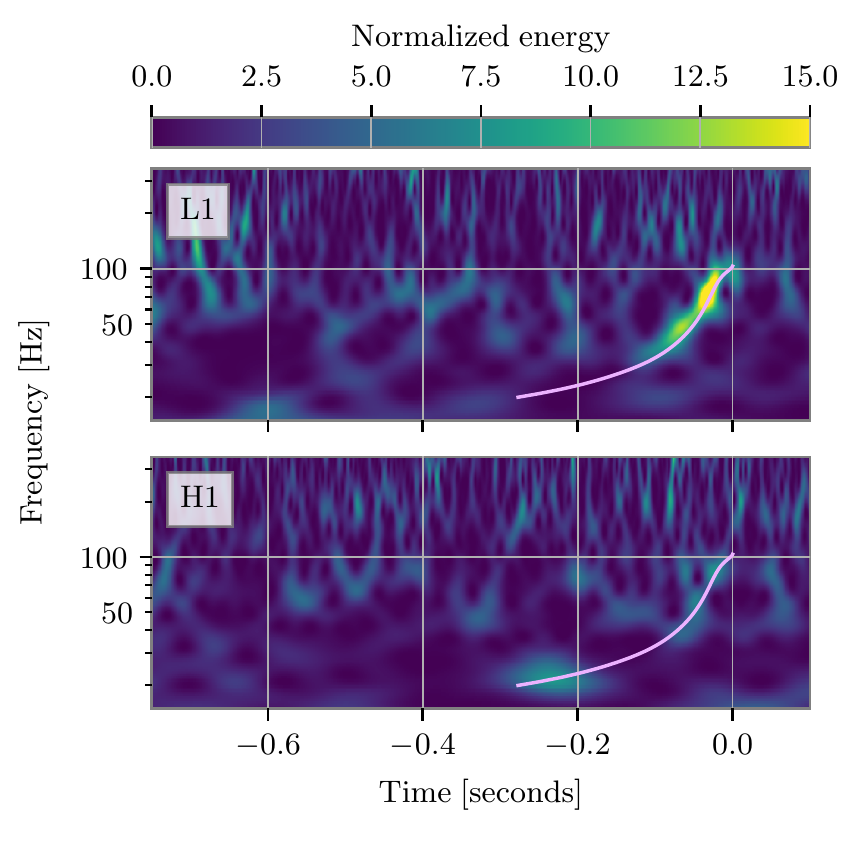}
    \caption{Time-frequency spectrograms of the \ofour event. We overlay the predicted signal using the \texttt{SEOBNRv4\_ROM} waveform approximant \citep{bohe_2017} and the median-estimated signal properties from our \pvtwo re-analysis of the event. (We note a mismatch between the waveform used for analysis and the waveform used to plot the signal track. However, the signal track here is intended only to guide the eye and not a quantitative test.)}
    \label{fig:omega_track}
\end{figure}

In \cref{fig:omega_track}, we plot the time-frequency spectrogram of \ofour and a characteristic signal track.
This illustrates that the event is primarily identified in the L1 detector, with no clear signal found in the H1 detector. 

\ofour was first identified in the GWTC2.1 catalogue by the GWTC2.1 \pycbc-BBH search alone \citep{GWTC-2-1_2021} with a probability of astrophysical origin (\pastro) of 0.61.
No other pipelines identified \ofour as a candidate.
Meanwhile, \onine was identified in the GWTC2 catalogue by the \gstlal search alone \citep{GWTC2_2021} with \pastro=1.
However, in the GWTC2.1 catalogue, \onine was subsequently identified by all four pipelines used with \pastro values ranging from 0.14 to 0.87.
\ofour is notable because the mass of the primary (heavier) object has more than 50\% support for a mass greater than \SI{100}{\Msun} and is either inside or above the ``mass gap'' predicted by pair-instability supernova theory \citep{woosley_2017}.

That \ofour and \onine are consistent with the Tomte glitch population does not, of course, mean that the events are Tomte glitches.
However, they have \pastro values, which allow for a terrestrial cause (i.e. $1-\pastro \gtrsim 0.1$). Tomte glitches were numerous during O3a and are known to impact search sensitivity adversely.
Therefore, it seems reasonable to conclude that if \ofour or \onine are not of astrophysical origin, of the four glitch classes considered here, they are most consistent with Tomte glitches.
However, we add that this assumes the signal morphology of the glitch classes studied herein extends to the low signal to noise ratio regime of \ofour and \onine.

This work does not change the estimated \pastro values for \ofour or \onine. The search pipelines that identify signals are robust to all relevant glitch classes, use information about the consistency between separate detectors, and apply glitch discriminators \citep{allen_2005}.
Using bootstrap approaches (e.g. time-shifting the data from independent detectors), the search pipelines estimate the background noise from all glitch classes, and this is intrinsically part of the \pastro calculation.
Nevertheless, \cref{fig:gwtc2_q_a1} does demonstrate that to identifying signals which have large spins and unequal mass ratios, astrophysical searches must contend with an elevated background of glitches compared to other areas of the parameter space.
Search pipelines account for the distribution of noise triggers (including glitches) over the searched parameter space in their detection statistics and the resulting significance estimates (both the False Alarm Rate, FAR, and \pastro), though this is done in a variety of ways \citep{nitz_2019, sachdev_2019, nitz_2020, aubin_2021}.
Therefore, the variation in glitch rate across the parameter space indicates that search pipelines are likely not uniformly sensitive across the parameter space \citep{davis_2020, abbott_2018_dq}.

\section{Discussion and Conclusion}
\label{sec:conclusion}
We present population models for the behaviour of non-Gaussian transient noise (glitches) in the Hanford and Livingston LIGO detectors during the O3a observing run.
We first identify lists of Blip, Tomte, Fast Scattering, and Scattered light glitches pre-categorised by the GravitySpy citizen-science project and then vetted for class consistency using time-frequency visualisations.
These glitch classes are known to most adversely search sensitivity.
Using Bayesian inference, we then analyse each glitch class using a precessing binary black hole model.
In effect, this allows us to project the properties of the glitch onto the parameter space of astrophysical signals.
Finally, we use hierarchical population inference methods to fit simple parameterised models to the glitch populations.
We do this separately for the chirp mass, mass ratio, and primary spin magnitude parameters; all other parameters we find to be only weakly informative at the population level.
We verify the performance of the models using posterior predictive checks and confirm that they reproduce the bulk features of the population.

Finally, we compare the population of glitches with events observed with $\pastro > 0.5$ during the O3a observing run.
We find two events, \ofour and \onine, which have a chirp mass, mass ratio, and primary spin magnitude consistent with the Tomte glitch class.
\ofour is a fascinating case as it has a mass that may be inconsistent with the predictions of pair-instability supernova theory.

The parameterised models developed in this work could be used by search pipelines to potentially better distinguish astrophysical signals from terrestrial glitches \footnote{For current pipelines, this would require a mapping between the precessing spin parameters used in this work to the aligned spin components used by search pipelines}.
This could be implemented either additively in the detection statistic used by matched-filtering pipelines or as part of the \emph{astrophysical odds} \citep{ashton_2019b} (see \citep{ashton_2020} for the use of a simplistic glitch model).
Already, many search pipelines use a version of this in which the detection statistics and background distributions are calculated over the template banks.
This enables a lowering of the background rate in regions of parameter space less contaminated by glitches.
Including the probabilistic models here into the detection statistic itself could refine this approach, providing more fine-grained modelled information.
Ultimately, this may lead to a greater number of astrophysical signals being detected.
However, the impact of this will depend on how much additional information is provided by these glitch models relative to the approaches already taken.
To determine if such an approach can improve the detection efficiency, a large scale simulation study is needed, which is beyond the scope of the current work.

We also identify multimodality in the mass ratio of Blip glitches which may be used to identify this glitch class.
We do not yet understand why this multimodality is exhibited, but it seems likely the waveform fits two separate parts of the glitch morphology in different regions of parameter space.
This information may help formulate an improved glitch model (e.g., breaking the astrophysical waveform in a non-physical way that fits the glitch).
Such a model could then be used to identify glitches and improve astrophysical searches.

At the same time, In the future, we hope to improve these models by including correlations between parameters, using higher-order mode waveform models and extending the study to sub-dominant parameters.
These improvements will provide a more refined glitch consistency test and may reveal new glitch population features.

\section{Acknowledgements}
We a grateful to Derek Davis for help with the script used to produce \cref{fig:glitch_examples} and \cref{fig:omega_track}, to Andrew Williamson for insightful comments about the nature of the bimodality seen in the Blip glitch mass ratio, to Tom Dent for helpful feedback on the manuscript, and to the CBC and detector characterisation groups of the LIGO and Virgo Collaborations for helpful feedback on this work. We also think Richard O'Shaughnessy, Jason Hathaway, Ben Henderson, Brian Cowburn for useful discussions related to this work. GA and LKN thank the UKRI Future Leaders Fellowship for support through the
grant MR/T01881X/1.
This research has made use of data, software and/or web tools obtained from the Gravitational Wave Open Science Center (https://www.gw-openscience.org), a service of LIGO Laboratory, the LIGO Scientific Collaboration and the Virgo Collaboration. LIGO is funded by the U.S. National Science Foundation. Virgo is funded by the French Centre National de Recherche Scientifique (CNRS), the Italian Istituto Nazionale della Fisica Nucleare (INFN) and the Dutch Nikhef, with contributions by Polish and Hungarian institutes.
We are also grateful to computing resources provided by the LIGO Laboratory computing clusters at California Institute of Technology and LIGO Hanford Observatory supported by National Science Foundation Grants PHY-0757058 and PHY-0823459.
The majority of analysis performed for this research was done using resources provided by the Open Science Grid \citep{osg07, osg09}, which is supported by the National Science Foundation award \#2030508.
 This work makes use of the
\texttt{scipy} \citep{scipy_2020},
\texttt{numpy} \citep{oliphant_2006, van_2011, harris_2020},
\texttt{gwpy} \citep{macleod_2021}, and
\texttt{PyCBC} \citep{nitz_2017} software for data analysis and visualisation.

\bibliography{bibliography}

\newcommand{\HanfordfastscatteringskewnormalmuZerochirpmassmedian}{164}
\newcommand{\HanfordfastscatteringskewnormalmuZerochirpmassminus}{8}
\newcommand{\HanfordfastscatteringskewnormalmuZerochirpmassplus}{7}
\newcommand{\HanfordfastscatteringskewnormalsigmaZerochirpmassmedian}{39}
\newcommand{\HanfordfastscatteringskewnormalsigmaZerochirpmassminus}{6}
\newcommand{\HanfordfastscatteringskewnormalsigmaZerochirpmassplus}{8}
\newcommand{\HanfordfastscatteringskewnormalkappaZerochirpmassmedian}{-7}
\newcommand{\HanfordfastscatteringskewnormalkappaZerochirpmassminus}{10}
\newcommand{\HanfordfastscatteringskewnormalkappaZerochirpmassplus}{5}
\newcommand{\HanfordfastscatteringskewnormalmuZeromassratiomedian}{0.120}
\newcommand{\HanfordfastscatteringskewnormalmuZeromassratiominus}{0.009}
\newcommand{\HanfordfastscatteringskewnormalmuZeromassratioplus}{0.005}
\newcommand{\HanfordfastscatteringskewnormalsigmaZeromassratiomedian}{0.027}
\newcommand{\HanfordfastscatteringskewnormalsigmaZeromassratiominus}{0.006}
\newcommand{\HanfordfastscatteringskewnormalsigmaZeromassratioplus}{0.006}
\newcommand{\HanfordfastscatteringskewnormalkappaZeromassratiomedian}{-7}
\newcommand{\HanfordfastscatteringskewnormalkappaZeromassratiominus}{10}
\newcommand{\HanfordfastscatteringskewnormalkappaZeromassratioplus}{6}
\newcommand{\HanfordfastscatteringpowerlawalphaZeroaOnemedian}{11}
\newcommand{\HanfordfastscatteringpowerlawalphaZeroaOneminus}{2}
\newcommand{\HanfordfastscatteringpowerlawalphaZeroaOneplus}{3}
\newcommand{\LivingstonscatteredlightpowerlawalphaZeroaOnemedian}{27}
\newcommand{\LivingstonscatteredlightpowerlawalphaZeroaOneminus}{5}
\newcommand{\LivingstonscatteredlightpowerlawalphaZeroaOneplus}{6}
\newcommand{\LivingstonscatteredlightskewnormalmuZerochirpmassmedian}{199}
\newcommand{\LivingstonscatteredlightskewnormalmuZerochirpmassminus}{10}
\newcommand{\LivingstonscatteredlightskewnormalmuZerochirpmassplus}{1}
\newcommand{\LivingstonscatteredlightskewnormalsigmaZerochirpmassmedian}{67}
\newcommand{\LivingstonscatteredlightskewnormalsigmaZerochirpmassminus}{9}
\newcommand{\LivingstonscatteredlightskewnormalsigmaZerochirpmassplus}{9}
\newcommand{\LivingstonscatteredlightskewnormalkappaZerochirpmassmedian}{-11}
\newcommand{\LivingstonscatteredlightskewnormalkappaZerochirpmassminus}{10}
\newcommand{\LivingstonscatteredlightskewnormalkappaZerochirpmassplus}{8}
\newcommand{\LivingstonscatteredlightskewnormalmuZeromassratiomedian}{0.116}
\newcommand{\LivingstonscatteredlightskewnormalmuZeromassratiominus}{0.05}
\newcommand{\LivingstonscatteredlightskewnormalmuZeromassratioplus}{0.007}
\newcommand{\LivingstonscatteredlightskewnormalsigmaZeromassratiomedian}{0.035}
\newcommand{\LivingstonscatteredlightskewnormalsigmaZeromassratiominus}{0.01}
\newcommand{\LivingstonscatteredlightskewnormalsigmaZeromassratioplus}{0.008}
\newcommand{\LivingstonscatteredlightskewnormalkappaZeromassratiomedian}{-4}
\newcommand{\LivingstonscatteredlightskewnormalkappaZeromassratiominus}{10}
\newcommand{\LivingstonscatteredlightskewnormalkappaZeromassratioplus}{6}
\newcommand{\HanfordscatteredlightskewnormalmuZerochirpmassmedian}{151}
\newcommand{\HanfordscatteredlightskewnormalmuZerochirpmassminus}{30}
\newcommand{\HanfordscatteredlightskewnormalmuZerochirpmassplus}{10}
\newcommand{\HanfordscatteredlightskewnormalsigmaZerochirpmassmedian}{26}
\newcommand{\HanfordscatteredlightskewnormalsigmaZerochirpmassminus}{7}
\newcommand{\HanfordscatteredlightskewnormalsigmaZerochirpmassplus}{9}
\newcommand{\HanfordscatteredlightskewnormalkappaZerochirpmassmedian}{-1}
\newcommand{\HanfordscatteredlightskewnormalkappaZerochirpmassminus}{2}
\newcommand{\HanfordscatteredlightskewnormalkappaZerochirpmassplus}{2}
\newcommand{\HanfordscatteredlightskewnormalmuZeromassratiomedian}{0.113}
\newcommand{\HanfordscatteredlightskewnormalmuZeromassratiominus}{0.006}
\newcommand{\HanfordscatteredlightskewnormalmuZeromassratioplus}{0.005}
\newcommand{\HanfordscatteredlightskewnormalsigmaZeromassratiomedian}{0.029}
\newcommand{\HanfordscatteredlightskewnormalsigmaZeromassratiominus}{0.005}
\newcommand{\HanfordscatteredlightskewnormalsigmaZeromassratioplus}{0.005}
\newcommand{\HanfordscatteredlightskewnormalkappaZeromassratiomedian}{-8}
\newcommand{\HanfordscatteredlightskewnormalkappaZeromassratiominus}{10}
\newcommand{\HanfordscatteredlightskewnormalkappaZeromassratioplus}{5}
\newcommand{\HanfordscatteredlightpowerlawalphaZeroaOnemedian}{29}
\newcommand{\HanfordscatteredlightpowerlawalphaZeroaOneminus}{5}
\newcommand{\HanfordscatteredlightpowerlawalphaZeroaOneplus}{5}
\newcommand{\HanfordBlipsskewnormalmuZerochirpmassmedian}{17.2}
\newcommand{\HanfordBlipsskewnormalmuZerochirpmassminus}{0.4}
\newcommand{\HanfordBlipsskewnormalmuZerochirpmassplus}{0.5}
\newcommand{\HanfordBlipsskewnormalsigmaZerochirpmassmedian}{11.8}
\newcommand{\HanfordBlipsskewnormalsigmaZerochirpmassminus}{0.6}
\newcommand{\HanfordBlipsskewnormalsigmaZerochirpmassplus}{0.6}
\newcommand{\HanfordBlipsskewnormalkappaZerochirpmassmedian}{4.1}
\newcommand{\HanfordBlipsskewnormalkappaZerochirpmassminus}{0.6}
\newcommand{\HanfordBlipsskewnormalkappaZerochirpmassplus}{0.7}
\newcommand{\HanfordBlipsnormalmuZeromassratiomedian}{0.104}
\newcommand{\HanfordBlipsnormalsigmaZeromassratiomedian}{0.030}
\newcommand{\HanfordBlipsnormalmuOnemassratiomedian}{0.299}
\newcommand{\HanfordBlipsnormalsigmaOnemassratiomedian}{0.018}
\newcommand{\HanfordBlipsximixturemassratiomedian}{0.20}
\newcommand{\HanfordBlipsnormalmuZeromassratiominus}{0.005}
\newcommand{\HanfordBlipsnormalmuZeromassratioplus}{0.005}
\newcommand{\HanfordBlipsnormalsigmaZeromassratiominus}{0.003}
\newcommand{\HanfordBlipsnormalsigmaZeromassratioplus}{0.005}
\newcommand{\HanfordBlipsnormalmuOnemassratiominus}{0.002}
\newcommand{\HanfordBlipsnormalmuOnemassratioplus}{0.002}
\newcommand{\HanfordBlipsnormalsigmaOnemassratiominus}{0.001}
\newcommand{\HanfordBlipsnormalsigmaOnemassratioplus}{0.002}
\newcommand{\HanfordBlipsximixturemassratiominus}{0.02}
\newcommand{\HanfordBlipsximixturemassratioplus}{0.03}
\newcommand{\HanfordBlipspowerlawalphaZeroaOnemedian}{58}
\newcommand{\HanfordBlipspowerlawalphaZeroaOneminus}{5}
\newcommand{\HanfordBlipspowerlawalphaZeroaOneplus}{4}
\newcommand{\LivingstonBlipsskewnormalmuZerochirpmassmedian}{16.1}
\newcommand{\LivingstonBlipsskewnormalmuZerochirpmassminus}{0.6}
\newcommand{\LivingstonBlipsskewnormalmuZerochirpmassplus}{0.6}
\newcommand{\LivingstonBlipsskewnormalsigmaZerochirpmassmedian}{14.7}
\newcommand{\LivingstonBlipsskewnormalsigmaZerochirpmassminus}{0.8}
\newcommand{\LivingstonBlipsskewnormalsigmaZerochirpmassplus}{0.9}
\newcommand{\LivingstonBlipsskewnormalkappaZerochirpmassmedian}{4.3}
\newcommand{\LivingstonBlipsskewnormalkappaZerochirpmassminus}{0.7}
\newcommand{\LivingstonBlipsskewnormalkappaZerochirpmassplus}{0.8}
\newcommand{\LivingstonBlipsnormalmuZeromassratiomedian}{0.101}
\newcommand{\LivingstonBlipsnormalmuZeromassratiominus}{0.01}
\newcommand{\LivingstonBlipsnormalmuZeromassratioplus}{0.006}
\newcommand{\LivingstonBlipsnormalsigmaZeromassratiomedian}{0.038}
\newcommand{\LivingstonBlipsnormalsigmaZeromassratiominus}{0.005}
\newcommand{\LivingstonBlipsnormalsigmaZeromassratioplus}{0.008}
\newcommand{\LivingstonBlipsnormalmuOnemassratiomedian}{0.317}
\newcommand{\LivingstonBlipsnormalmuOnemassratiominus}{0.002}
\newcommand{\LivingstonBlipsnormalmuOnemassratioplus}{0.003}
\newcommand{\LivingstonBlipsnormalsigmaOnemassratiomedian}{0.025}
\newcommand{\LivingstonBlipsnormalsigmaOnemassratiominus}{0.002}
\newcommand{\LivingstonBlipsnormalsigmaOnemassratioplus}{0.002}
\newcommand{\LivingstonBlipsximixturemassratiomedian}{0.21}
\newcommand{\LivingstonBlipsximixturemassratiominus}{0.03}
\newcommand{\LivingstonBlipsximixturemassratioplus}{0.03}
\newcommand{\LivingstonBlipspowerlawalphaZeroaOnemedian}{60}
\newcommand{\LivingstonBlipspowerlawalphaZeroaOneminus}{4}
\newcommand{\LivingstonBlipspowerlawalphaZeroaOneplus}{5}
\newcommand{\LivingstonfastscatteringskewnormalmuZerochirpmassmedian}{102}
\newcommand{\LivingstonfastscatteringskewnormalmuZerochirpmassminus}{8}
\newcommand{\LivingstonfastscatteringskewnormalmuZerochirpmassplus}{30}
\newcommand{\LivingstonfastscatteringskewnormalsigmaZerochirpmassmedian}{43}
\newcommand{\LivingstonfastscatteringskewnormalsigmaZerochirpmassminus}{20}
\newcommand{\LivingstonfastscatteringskewnormalsigmaZerochirpmassplus}{10}
\newcommand{\LivingstonfastscatteringskewnormalkappaZerochirpmassmedian}{4}
\newcommand{\LivingstonfastscatteringskewnormalkappaZerochirpmassminus}{4}
\newcommand{\LivingstonfastscatteringskewnormalkappaZerochirpmassplus}{6}
\newcommand{\LivingstonfastscatteringskewnormalmuZeromassratiomedian}{0.072}
\newcommand{\LivingstonfastscatteringskewnormalmuZeromassratiominus}{0.006}
\newcommand{\LivingstonfastscatteringskewnormalmuZeromassratioplus}{0.006}
\newcommand{\LivingstonfastscatteringskewnormalsigmaZeromassratiomedian}{0.077}
\newcommand{\LivingstonfastscatteringskewnormalsigmaZeromassratiominus}{0.02}
\newcommand{\LivingstonfastscatteringskewnormalsigmaZeromassratioplus}{0.02}
\newcommand{\LivingstonfastscatteringskewnormalkappaZeromassratiomedian}{11}
\newcommand{\LivingstonfastscatteringskewnormalkappaZeromassratiominus}{5}
\newcommand{\LivingstonfastscatteringskewnormalkappaZeromassratioplus}{10}
\newcommand{\LivingstonfastscatteringpowerlawalphaZeroaOnemedian}{21}
\newcommand{\LivingstonfastscatteringpowerlawalphaZeroaOneminus}{4}
\newcommand{\LivingstonfastscatteringpowerlawalphaZeroaOneplus}{6}
\newcommand{\LivingstonTomtesskewnormalmuZerochirpmassmedian}{83.5}
\newcommand{\LivingstonTomtesskewnormalmuZerochirpmassminus}{0.6}
\newcommand{\LivingstonTomtesskewnormalmuZerochirpmassplus}{0.6}
\newcommand{\LivingstonTomtesskewnormalsigmaZerochirpmassmedian}{8.8}
\newcommand{\LivingstonTomtesskewnormalsigmaZerochirpmassminus}{0.6}
\newcommand{\LivingstonTomtesskewnormalsigmaZerochirpmassplus}{0.7}
\newcommand{\LivingstonTomtesskewnormalkappaZerochirpmassmedian}{-10}
\newcommand{\LivingstonTomtesskewnormalkappaZerochirpmassminus}{10}
\newcommand{\LivingstonTomtesskewnormalkappaZerochirpmassplus}{6}
\newcommand{\LivingstonTomtesskewnormalmuZeromassratiomedian}{0.283}
\newcommand{\LivingstonTomtesskewnormalmuZeromassratiominus}{0.003}
\newcommand{\LivingstonTomtesskewnormalmuZeromassratioplus}{0.003}
\newcommand{\LivingstonTomtesskewnormalsigmaZeromassratiomedian}{0.043}
\newcommand{\LivingstonTomtesskewnormalsigmaZeromassratiominus}{0.003}
\newcommand{\LivingstonTomtesskewnormalsigmaZeromassratioplus}{0.003}
\newcommand{\LivingstonTomtesskewnormalkappaZeromassratiomedian}{-16}
\newcommand{\LivingstonTomtesskewnormalkappaZeromassratiominus}{10}
\newcommand{\LivingstonTomtesskewnormalkappaZeromassratioplus}{7}
\newcommand{\LivingstonTomtespowerlawalphaZeroaOnemedian}{6.9}
\newcommand{\LivingstonTomtespowerlawalphaZeroaOneminus}{0.6}
\newcommand{\LivingstonTomtespowerlawalphaZeroaOneplus}{0.6}
\newcommand{\HanfordTomtesskewnormalmuZerochirpmassmedian}{56}
\newcommand{\HanfordTomtesskewnormalmuZerochirpmassminus}{1}
\newcommand{\HanfordTomtesskewnormalmuZerochirpmassplus}{1}
\newcommand{\HanfordTomtesskewnormalsigmaZerochirpmassmedian}{29}
\newcommand{\HanfordTomtesskewnormalsigmaZerochirpmassminus}{2}
\newcommand{\HanfordTomtesskewnormalsigmaZerochirpmassplus}{2}
\newcommand{\HanfordTomtesskewnormalkappaZerochirpmassmedian}{3.6}
\newcommand{\HanfordTomtesskewnormalkappaZerochirpmassminus}{0.6}
\newcommand{\HanfordTomtesskewnormalkappaZerochirpmassplus}{0.7}
\newcommand{\HanfordTomtesskewnormalmuZeromassratiomedian}{0.289}
\newcommand{\HanfordTomtesskewnormalmuZeromassratiominus}{0.003}
\newcommand{\HanfordTomtesskewnormalmuZeromassratioplus}{0.003}
\newcommand{\HanfordTomtesskewnormalsigmaZeromassratiomedian}{0.057}
\newcommand{\HanfordTomtesskewnormalsigmaZeromassratiominus}{0.003}
\newcommand{\HanfordTomtesskewnormalsigmaZeromassratioplus}{0.004}
\newcommand{\HanfordTomtesskewnormalkappaZeromassratiomedian}{-14}
\newcommand{\HanfordTomtesskewnormalkappaZeromassratiominus}{10}
\newcommand{\HanfordTomtesskewnormalkappaZeromassratioplus}{6}
\newcommand{\HanfordTomtespowerlawalphaZeroaOnemedian}{5.3}
\newcommand{\HanfordTomtespowerlawalphaZeroaOneminus}{0.4}
\newcommand{\HanfordTomtespowerlawalphaZeroaOneplus}{0.4}

\begin{table*}[h]
    \centering
    \begin{tabular}{c|c|c|c|l|c}
         Detector & Type & Parameter & Model & Median & Posterior Predictive  \\ \hline\hline
         H1 & Tomte & \mchirp & M2: Skew-Normal & \begin{tabular}{l}
         $\mu=\HanfordTomtesskewnormalmuZerochirpmassmedian_{-\HanfordTomtesskewnormalmuZerochirpmassminus}^{+\HanfordTomtesskewnormalmuZerochirpmassplus}$\\
$\sigma=\HanfordTomtesskewnormalsigmaZerochirpmassmedian_{-\HanfordTomtesskewnormalsigmaZerochirpmassminus}^{+\HanfordTomtesskewnormalsigmaZerochirpmassplus}$\\
$\kappa=\HanfordTomtesskewnormalkappaZerochirpmassmedian_{-\HanfordTomtesskewnormalkappaZerochirpmassminus}^{+\HanfordTomtesskewnormalkappaZerochirpmassplus}$\\
         \end{tabular} & 
         \ici{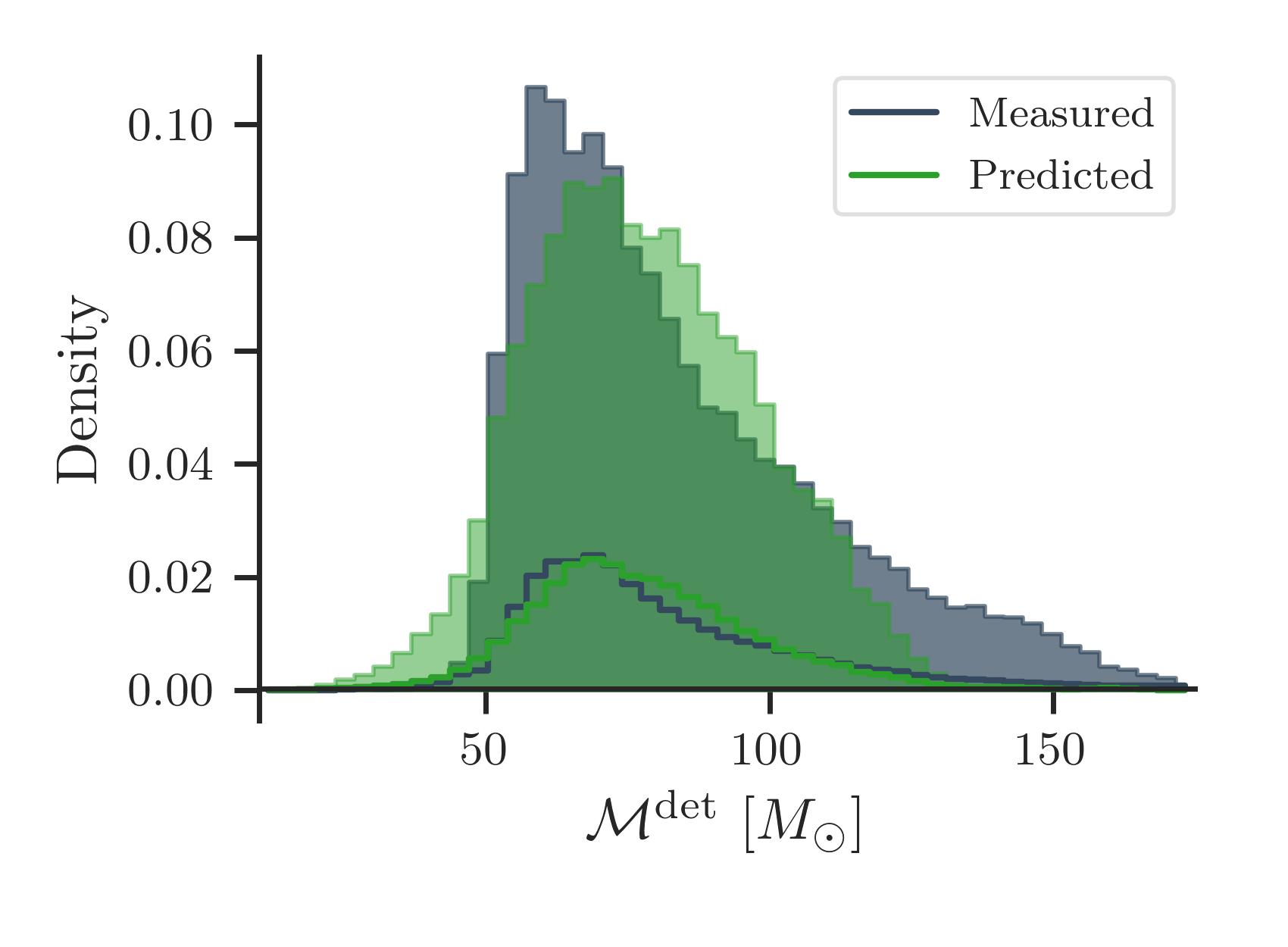}
         \\\hline
         L1 & Tomte & \mchirp & M2: Skew-Normal & \begin{tabular}{l}
$\mu=\LivingstonTomtesskewnormalmuZerochirpmassmedian_{-\LivingstonTomtesskewnormalmuZerochirpmassminus}^{+\LivingstonTomtesskewnormalmuZerochirpmassplus}$\\
$\sigma=\LivingstonTomtesskewnormalsigmaZerochirpmassmedian_{-\LivingstonTomtesskewnormalsigmaZerochirpmassminus}^{+\LivingstonTomtesskewnormalsigmaZerochirpmassplus}$\\
$\kappa=\LivingstonTomtesskewnormalkappaZerochirpmassmedian_{-\LivingstonTomtesskewnormalkappaZerochirpmassminus}^{+\LivingstonTomtesskewnormalkappaZerochirpmassplus}$\\
         \end{tabular} & 
         \ici{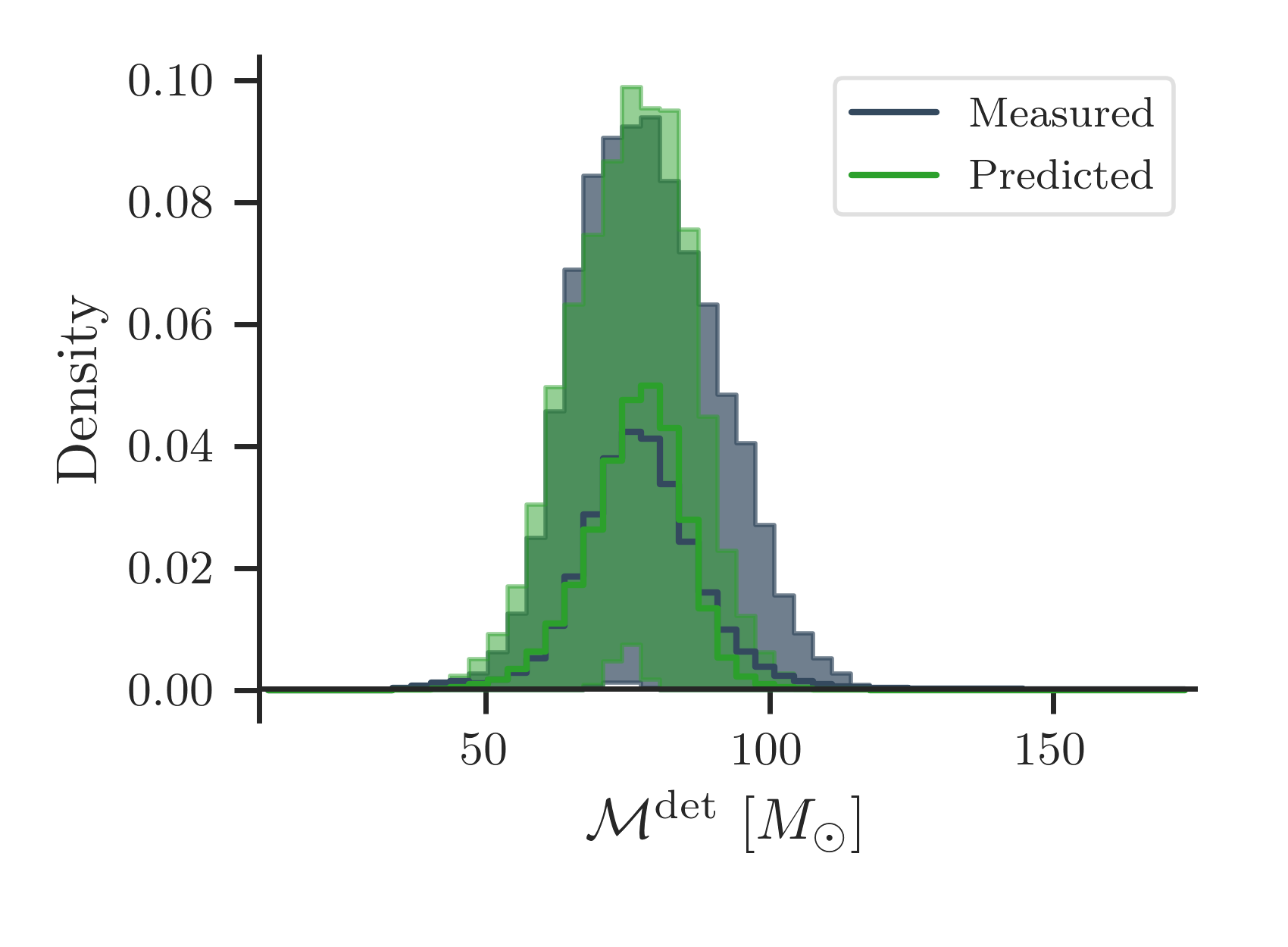}
         \\\hline
         H1 & Tomte & \massratio & M2: Skew-Normal & \begin{tabular}{l}
$\mu=\HanfordTomtesskewnormalmuZeromassratiomedian_{-\HanfordTomtesskewnormalmuZeromassratiominus}^{+\HanfordTomtesskewnormalmuZeromassratioplus}$\\
$\sigma=\HanfordTomtesskewnormalsigmaZeromassratiomedian_{-\HanfordTomtesskewnormalsigmaZeromassratiominus}^{+\HanfordTomtesskewnormalsigmaZeromassratioplus}$\\
$\kappa=\HanfordTomtesskewnormalkappaZeromassratiomedian_{-\HanfordTomtesskewnormalkappaZeromassratiominus}^{+\HanfordTomtesskewnormalkappaZeromassratioplus}$\\
         \end{tabular} &        
         \ici{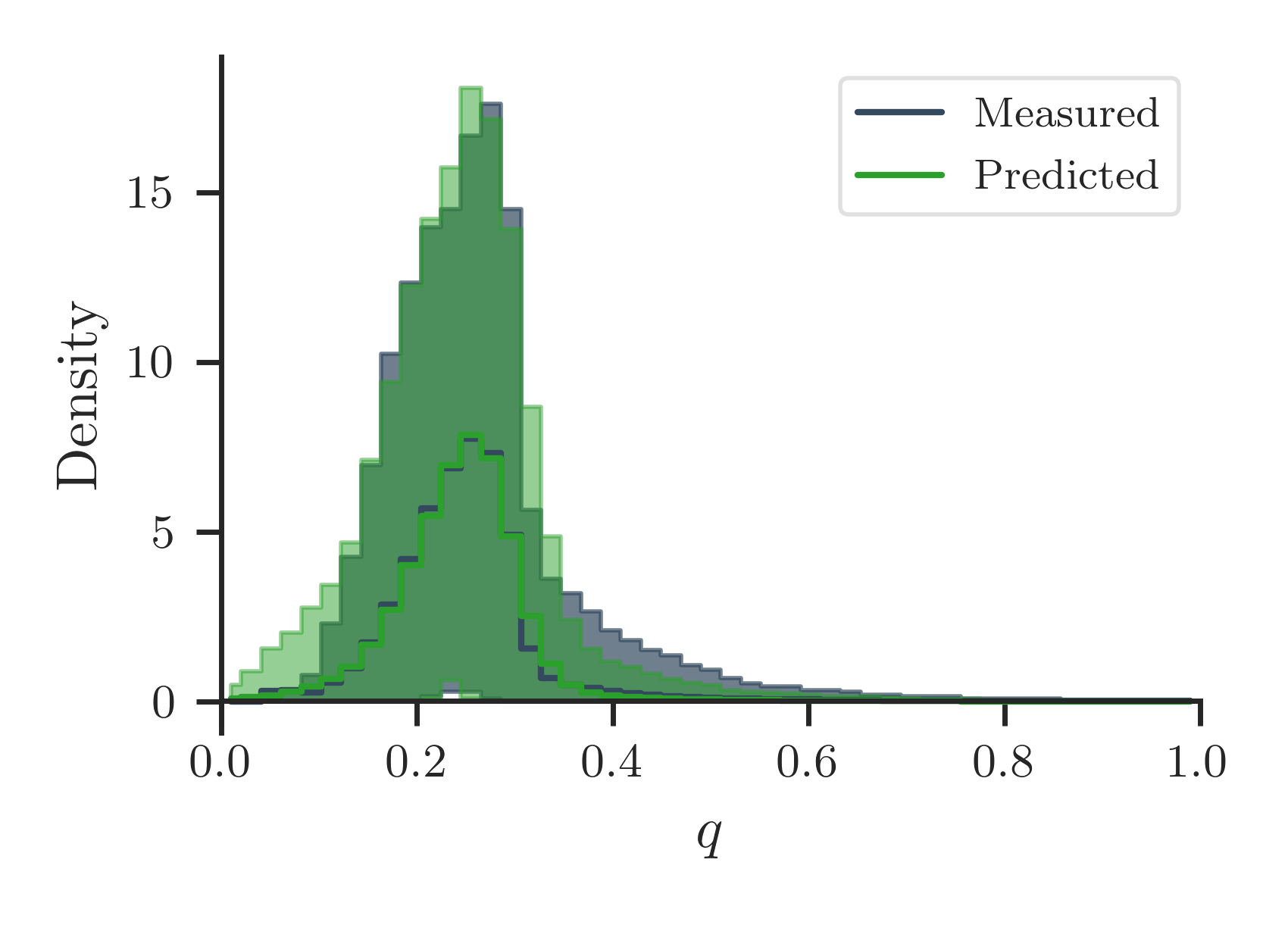}
         \\\hline
         L1 & Tomte & \massratio & M2: Skew-Normal & \begin{tabular}{c}
$\mu=\LivingstonTomtesskewnormalmuZeromassratiomedian_{-\LivingstonTomtesskewnormalmuZeromassratiominus}^{+\LivingstonTomtesskewnormalmuZeromassratioplus}$\\
$\sigma=\LivingstonTomtesskewnormalsigmaZeromassratiomedian_{-\LivingstonTomtesskewnormalsigmaZeromassratiominus}^{+\LivingstonTomtesskewnormalsigmaZeromassratioplus}$\\
$\kappa=\LivingstonTomtesskewnormalkappaZeromassratiomedian_{-\LivingstonTomtesskewnormalkappaZeromassratiominus}^{+\LivingstonTomtesskewnormalkappaZeromassratioplus}$\\
         \end{tabular} &        
         \ici{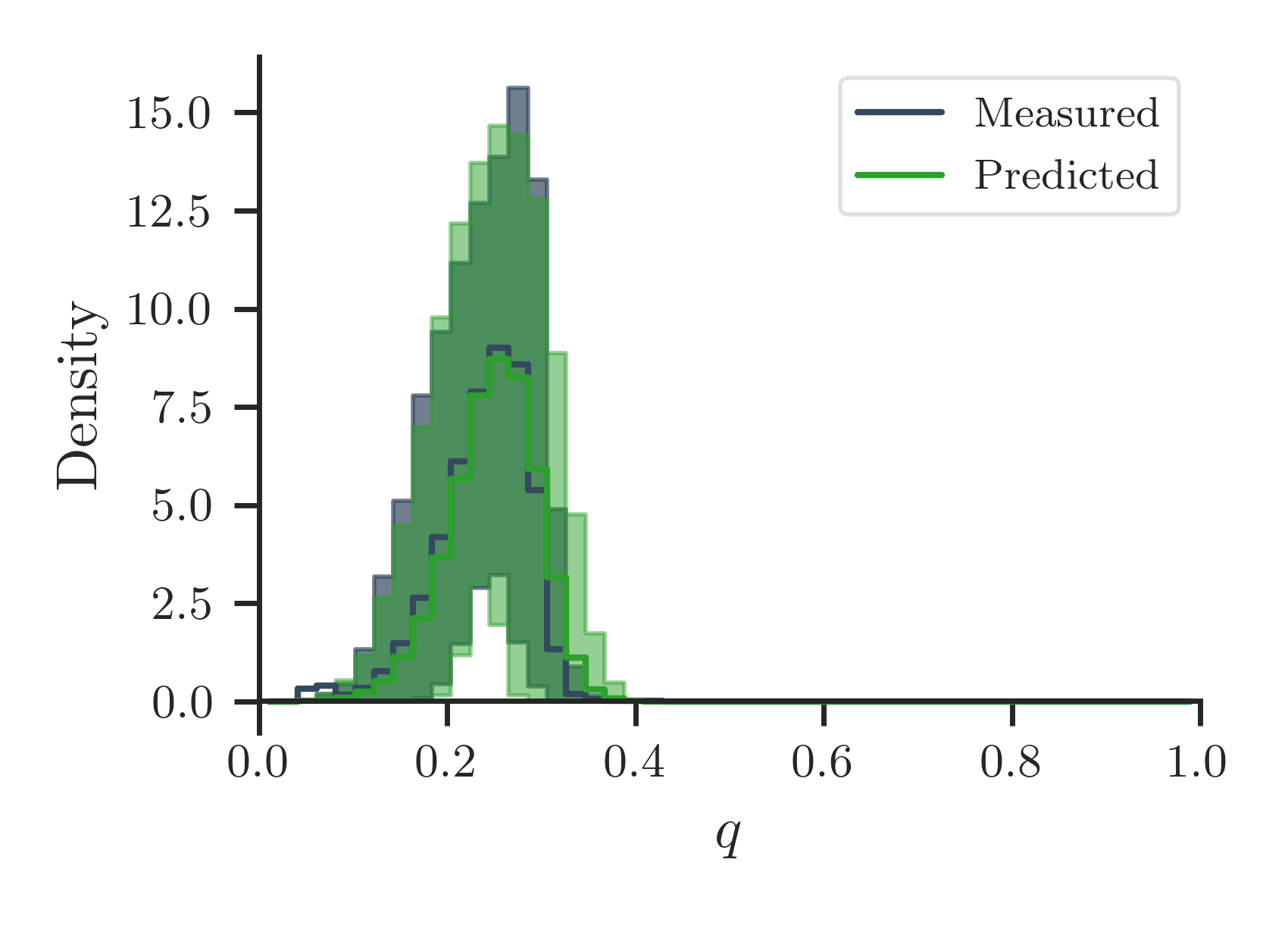}
         \\\hline
         H1 & Tomte & \primaryspin & M1: Powerlaw & \begin{tabular}{c}
$\alpha=\HanfordTomtespowerlawalphaZeroaOnemedian_{-\HanfordTomtespowerlawalphaZeroaOneminus}^{+\HanfordTomtespowerlawalphaZeroaOneplus}$\\
         \end{tabular} &        
         \ici{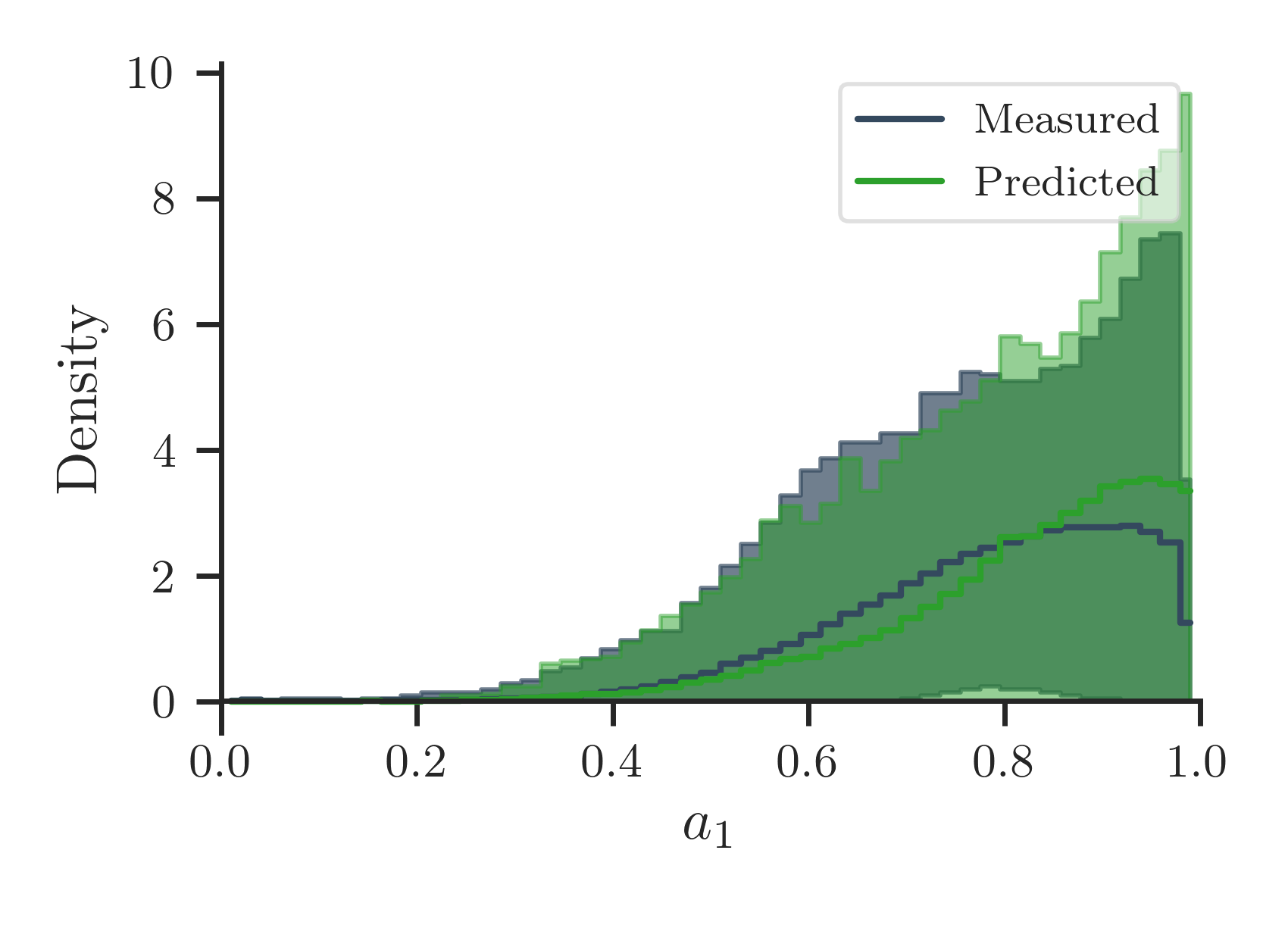}
         \\\hline
         L1 & Tomte & \primaryspin & M1: Powerlaw & \begin{tabular}{c}
$\alpha=\LivingstonTomtespowerlawalphaZeroaOnemedian_{-\LivingstonTomtespowerlawalphaZeroaOneminus}^{+\LivingstonTomtespowerlawalphaZeroaOneplus}$\\
         \end{tabular} &        
         \ici{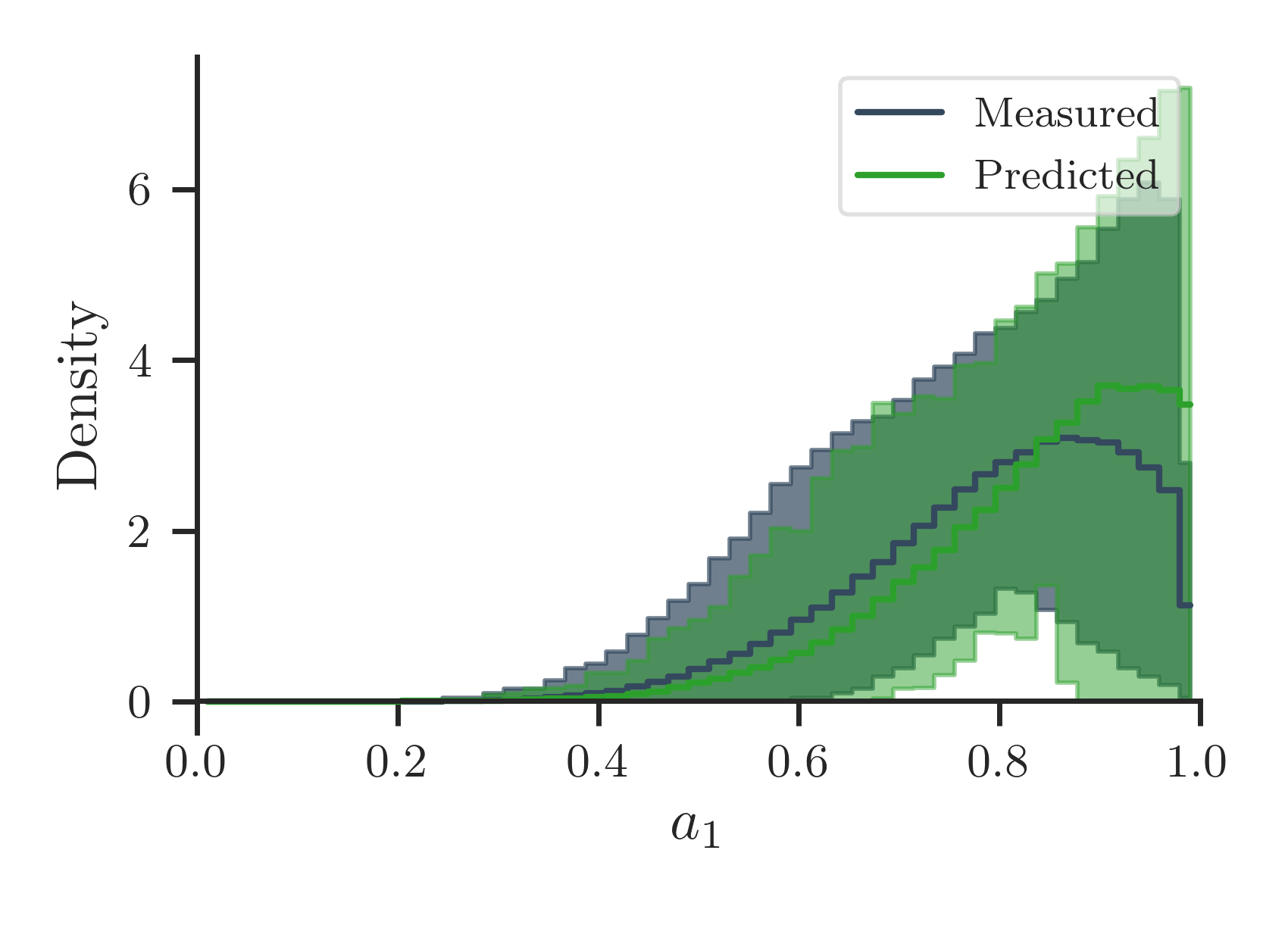}
         \\\hline
    \end{tabular}
    \caption{Table of parameterised models for Tomte glitches in the H1 and L1 interferometers during the O3a era. Uncertainties on the median denote the symmetric 90\% credible interval. See Sec.~\ref{sec:methodology} for a description of the posterior predictive plots.}
    \label{tab:tomte}
\end{table*}

\begin{table*}[h]
    \centering
    \begin{tabular}{c|c|c|c|l|c}
         Detector & Type & Parameter & Model & Median & Posterior Predictive  \\ \hline\hline
         H1 & Blip & \mchirp & M2: Skew-Normal & \begin{tabular}{l}
$\mu=\HanfordBlipsskewnormalmuZerochirpmassmedian_{-\HanfordBlipsskewnormalmuZerochirpmassminus}^{+\HanfordBlipsskewnormalmuZerochirpmassplus}$\\
$\sigma=\HanfordBlipsskewnormalsigmaZerochirpmassmedian_{-\HanfordBlipsskewnormalsigmaZerochirpmassminus}^{+\HanfordBlipsskewnormalsigmaZerochirpmassplus}$\\
$\kappa=\HanfordBlipsskewnormalkappaZerochirpmassmedian_{-\HanfordBlipsskewnormalkappaZerochirpmassminus}^{+\HanfordBlipsskewnormalkappaZerochirpmassplus}$\\
         \end{tabular} & 
         \ici{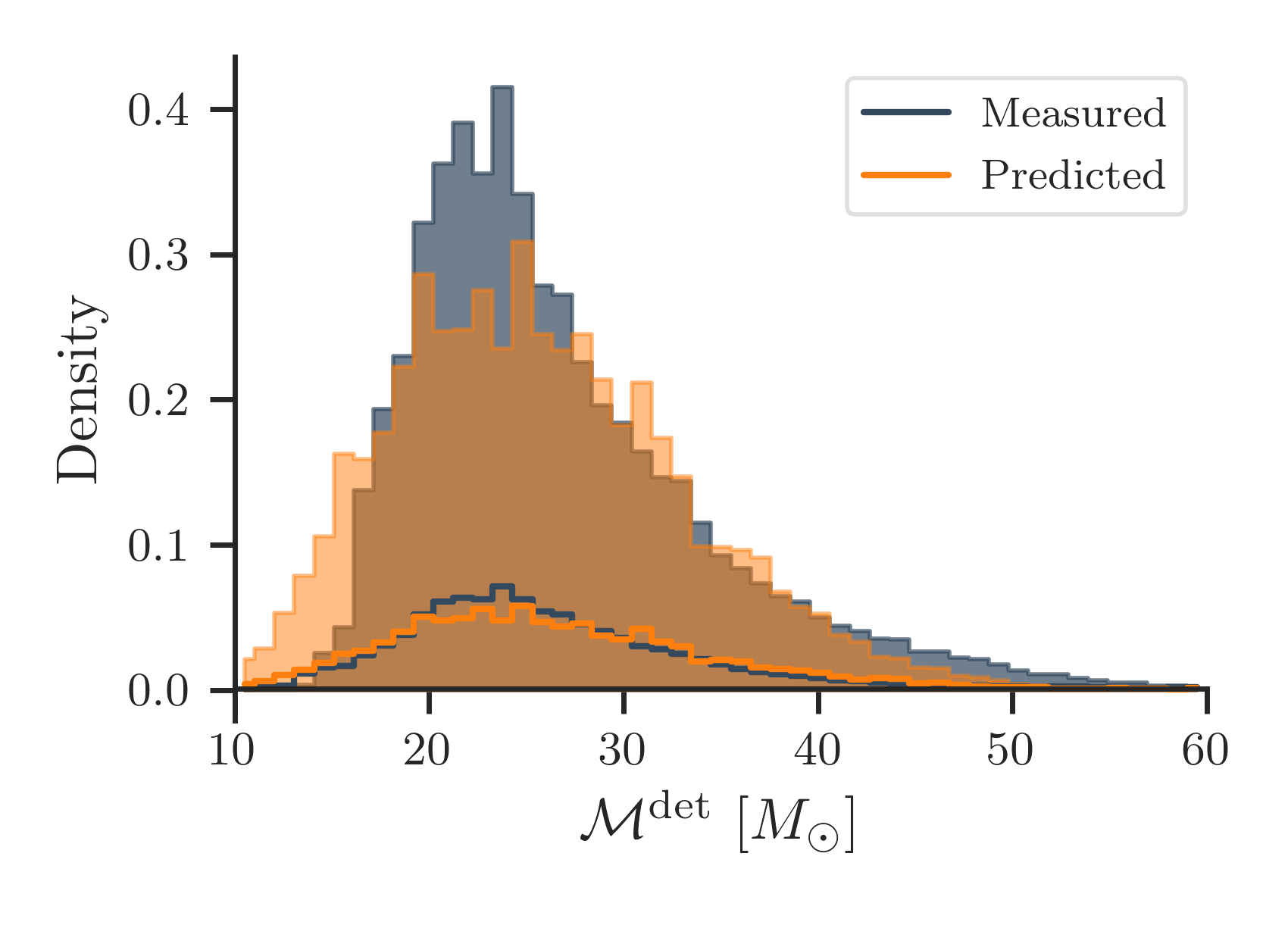}
         \\\hline
         L1 & Blip & \mchirp & M2: Skew-Normal & \begin{tabular}{l}
$\mu=\LivingstonBlipsskewnormalmuZerochirpmassmedian_{-\LivingstonBlipsskewnormalmuZerochirpmassminus}^{+\LivingstonBlipsskewnormalmuZerochirpmassplus}$\\
$\sigma=\LivingstonBlipsskewnormalsigmaZerochirpmassmedian_{-\LivingstonBlipsskewnormalsigmaZerochirpmassminus}^{+\LivingstonBlipsskewnormalsigmaZerochirpmassplus}$\\
$\kappa=\LivingstonBlipsskewnormalkappaZerochirpmassmedian_{-\LivingstonBlipsskewnormalkappaZerochirpmassminus}^{+\LivingstonBlipsskewnormalkappaZerochirpmassplus}$\\
         \end{tabular} & 
         \ici{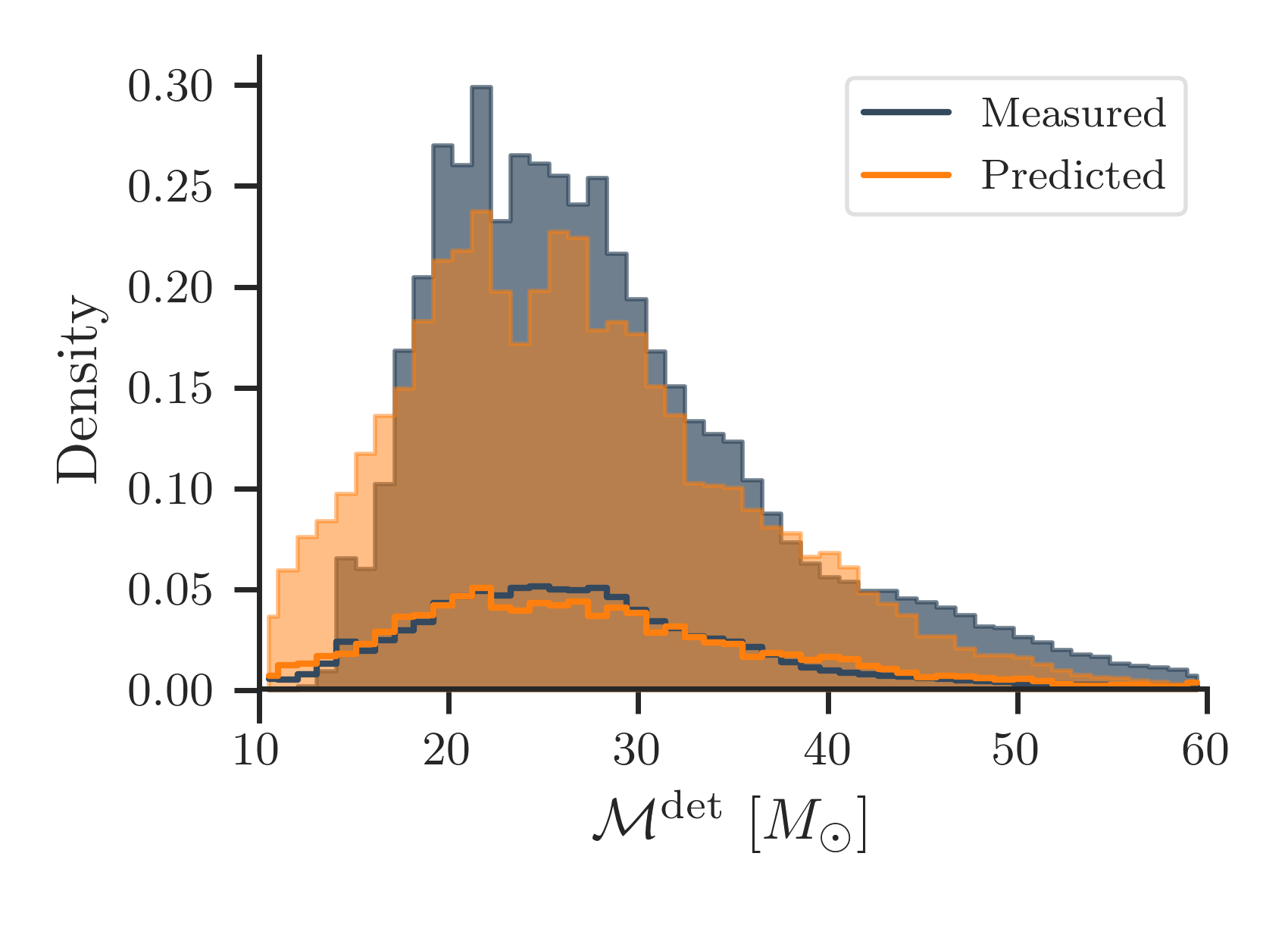}
         \\\hline
         H1 & Blip & \massratio & M3: Normal+Normal & \begin{tabular}{l}
$\mu_0=\HanfordBlipsnormalmuZeromassratiomedian_{-\HanfordBlipsnormalmuZeromassratiominus}^{+\HanfordBlipsnormalmuZeromassratioplus}$\\
$\sigma_0=\HanfordBlipsnormalsigmaZeromassratiomedian_{-\HanfordBlipsnormalsigmaZeromassratiominus}^{+\HanfordBlipsnormalsigmaZeromassratioplus}$\\
$\mu_1=\HanfordBlipsnormalmuOnemassratiomedian_{-\HanfordBlipsnormalmuOnemassratiominus}^{+\HanfordBlipsnormalmuOnemassratioplus}$\\
$\sigma_1=\HanfordBlipsnormalsigmaOnemassratiomedian_{-\HanfordBlipsnormalsigmaOnemassratiominus}^{+\HanfordBlipsnormalsigmaOnemassratioplus}$\\
$\xi=\HanfordBlipsximixturemassratiomedian_{-\HanfordBlipsximixturemassratiominus}^{+\HanfordBlipsximixturemassratioplus}$\\
         \end{tabular} &        
         \ici{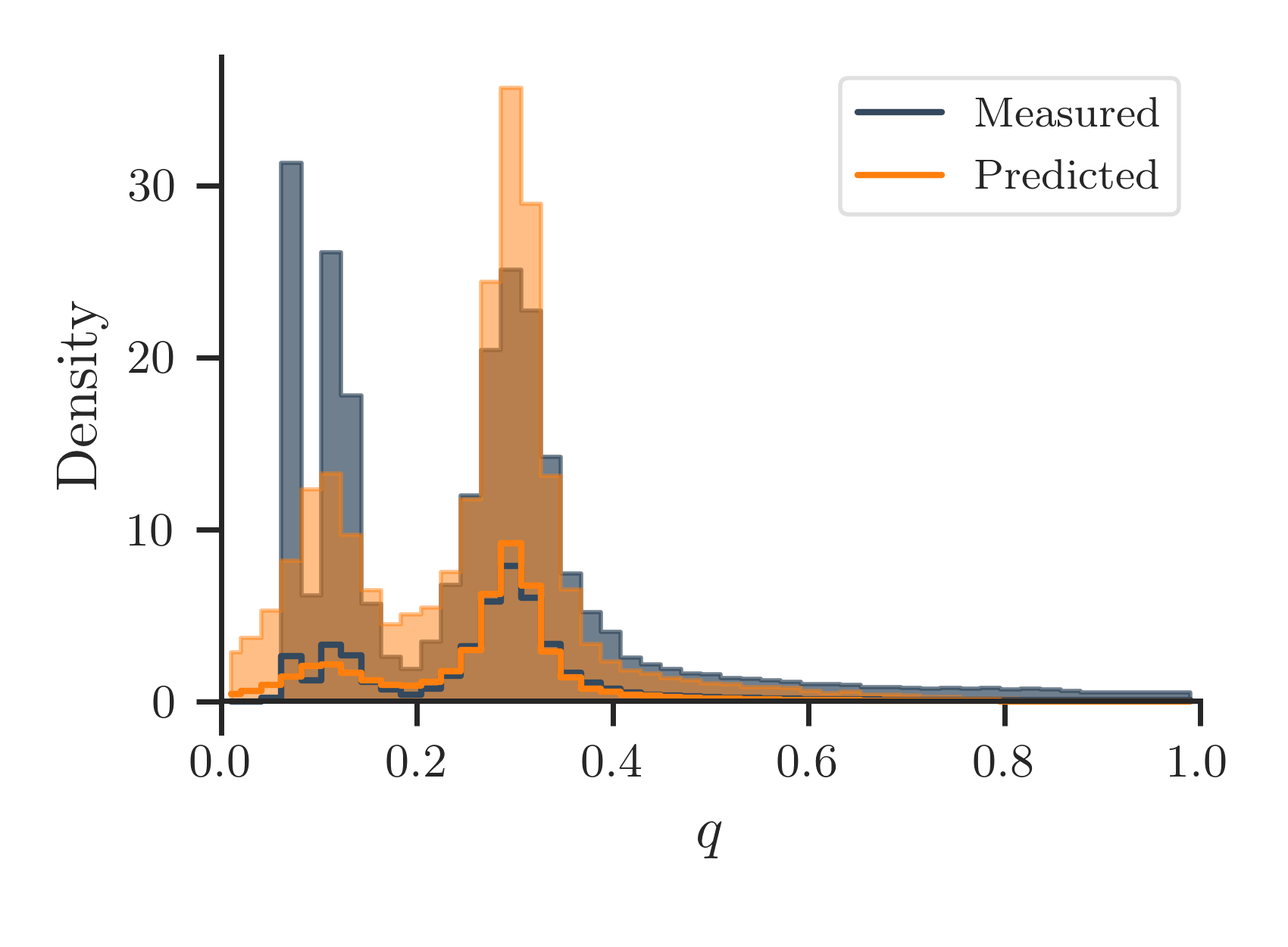}
         \\\hline
         L1 & Blip & \massratio & M3: Normal+Normal & \begin{tabular}{l}
$\mu_0=\LivingstonBlipsnormalmuZeromassratiomedian_{-\LivingstonBlipsnormalmuZeromassratiominus}^{+\LivingstonBlipsnormalmuZeromassratioplus}$\\
$\sigma_0=\LivingstonBlipsnormalsigmaZeromassratiomedian_{-\LivingstonBlipsnormalsigmaZeromassratiominus}^{+\LivingstonBlipsnormalsigmaZeromassratioplus}$\\
$\mu_1=\LivingstonBlipsnormalmuOnemassratiomedian_{-\LivingstonBlipsnormalmuOnemassratiominus}^{+\LivingstonBlipsnormalmuOnemassratioplus}$\\
$\sigma_1=\LivingstonBlipsnormalsigmaOnemassratiomedian_{-\LivingstonBlipsnormalsigmaOnemassratiominus}^{+\LivingstonBlipsnormalsigmaOnemassratioplus}$\\
$\xi=\LivingstonBlipsximixturemassratiomedian_{-\LivingstonBlipsximixturemassratiominus}^{+\LivingstonBlipsximixturemassratioplus}$\\
         \end{tabular} &        
         \ici{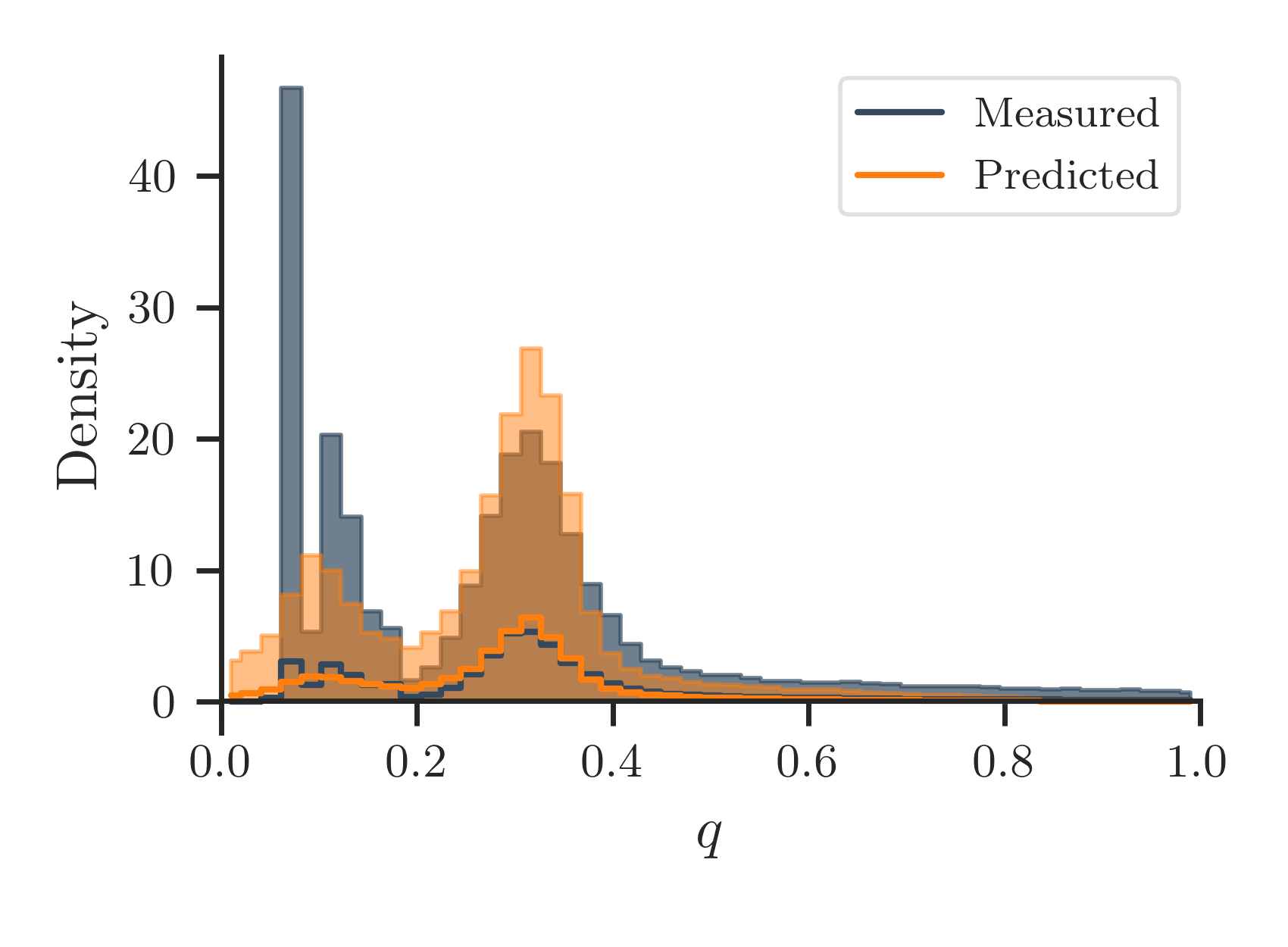}
         \\\hline
         H1 & Blip & \primaryspin & M1: Powerlaw & \begin{tabular}{l}
$\alpha=\HanfordBlipspowerlawalphaZeroaOnemedian_{-\HanfordBlipspowerlawalphaZeroaOneminus}^{+\HanfordBlipspowerlawalphaZeroaOneplus}$\\
         \end{tabular} &        
         \ici{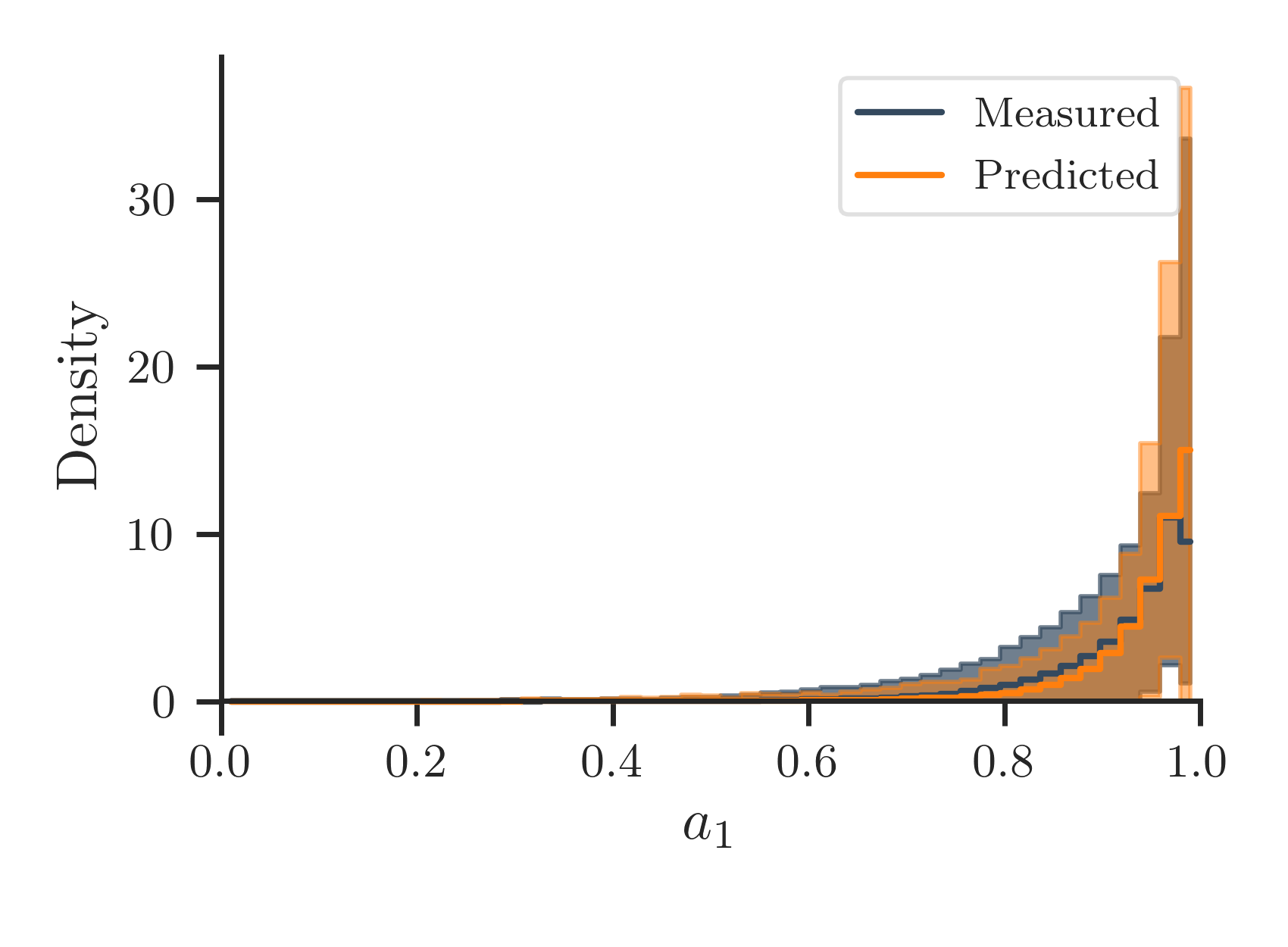}
         \\\hline
         L1 & Blip & \primaryspin & M1: Powerlaw & \begin{tabular}{l}
$\alpha=\LivingstonBlipspowerlawalphaZeroaOnemedian_{-\LivingstonBlipspowerlawalphaZeroaOneminus}^{+\LivingstonBlipspowerlawalphaZeroaOneplus}$\\
         \end{tabular} &        
         \ici{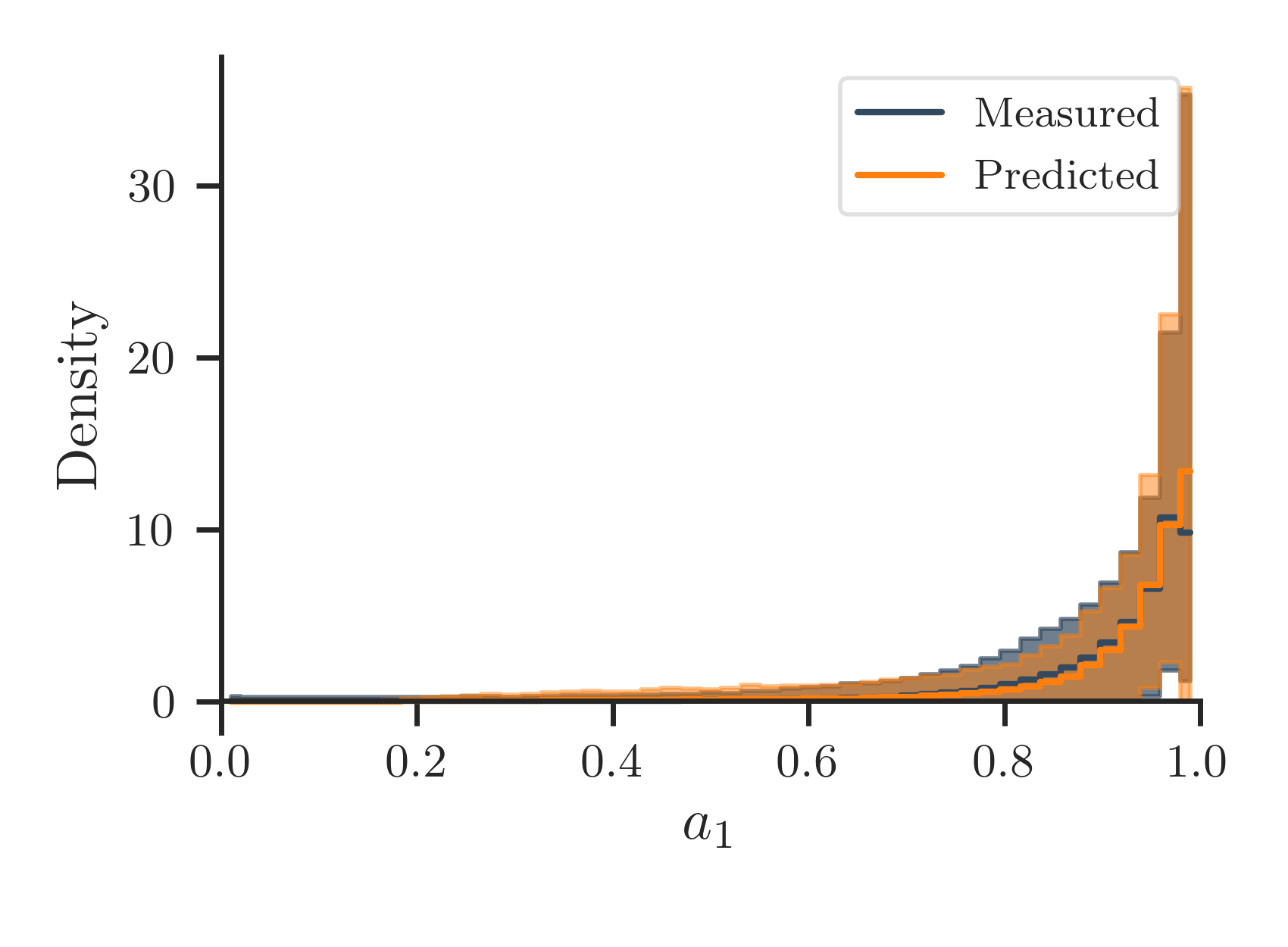}
         \\\hline
    \end{tabular}
    \caption{Table of parameterised models for Blip glitches in the H1 and L1 interferometers during the O3a era. Uncertainties on the median denote the symmetric 90\% credible interval. See Sec.~\ref{sec:methodology} for a description of the posterior predictive plots. See Sec.~\ref{sec:methodology} for a description of the posterior predictive plots.}
    \label{tab:blip}
\end{table*}

\begin{table*}[h]
    \centering
    \begin{tabular}{c|c|c|c|l|c}
         Detector & Type & Parameter & Model & Median & Posterior Predictive  \\ \hline\hline
         H1 & Fast Scattering & \mchirp & M2: Skew-Normal & \begin{tabular}{c}
$\mu=\HanfordfastscatteringskewnormalmuZerochirpmassmedian_{-\HanfordfastscatteringskewnormalmuZerochirpmassminus}^{+\HanfordfastscatteringskewnormalmuZerochirpmassplus}$\\
$\sigma=\HanfordfastscatteringskewnormalsigmaZerochirpmassmedian_{-\HanfordfastscatteringskewnormalsigmaZerochirpmassminus}^{+\HanfordfastscatteringskewnormalsigmaZerochirpmassplus}$\\
$\kappa=\HanfordfastscatteringskewnormalkappaZerochirpmassmedian_{-\HanfordfastscatteringskewnormalkappaZerochirpmassminus}^{+\HanfordfastscatteringskewnormalkappaZerochirpmassplus}$\\
         \end{tabular} & 
         \ici{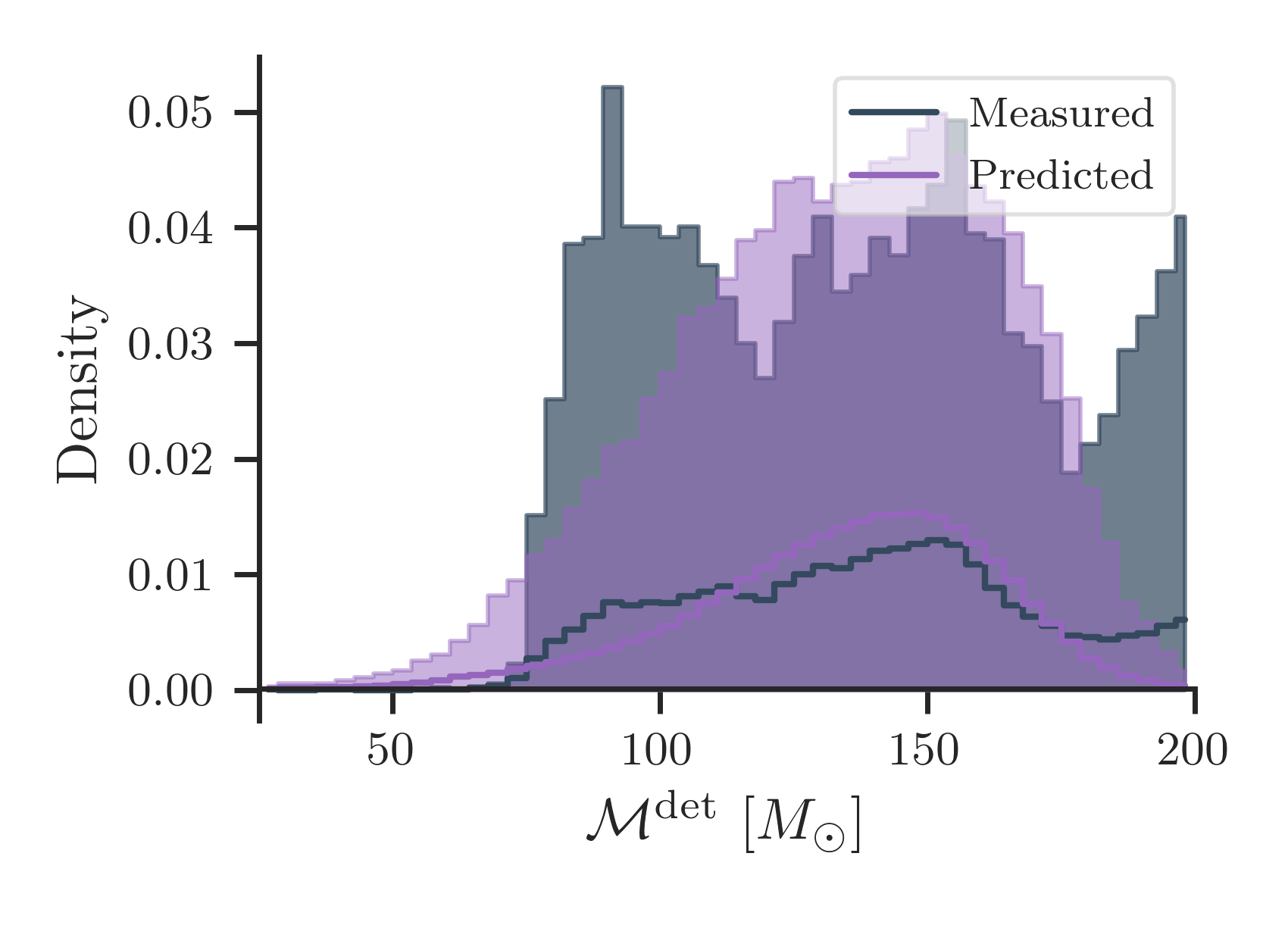}
         \\\hline
         L1 & Fast Scattering & \mchirp & M2: Skew-Normal & \begin{tabular}{c}
$\mu=\LivingstonfastscatteringskewnormalmuZerochirpmassmedian_{-\LivingstonfastscatteringskewnormalmuZerochirpmassminus}^{+\LivingstonfastscatteringskewnormalmuZerochirpmassplus}$\\
$\sigma=\LivingstonfastscatteringskewnormalsigmaZerochirpmassmedian_{-\LivingstonfastscatteringskewnormalsigmaZerochirpmassminus}^{+\LivingstonfastscatteringskewnormalsigmaZerochirpmassplus}$\\
$\kappa=\LivingstonfastscatteringskewnormalkappaZerochirpmassmedian_{-\LivingstonfastscatteringskewnormalkappaZerochirpmassminus}^{+\LivingstonfastscatteringskewnormalkappaZerochirpmassplus}$\\
         \end{tabular} & 
         \ici{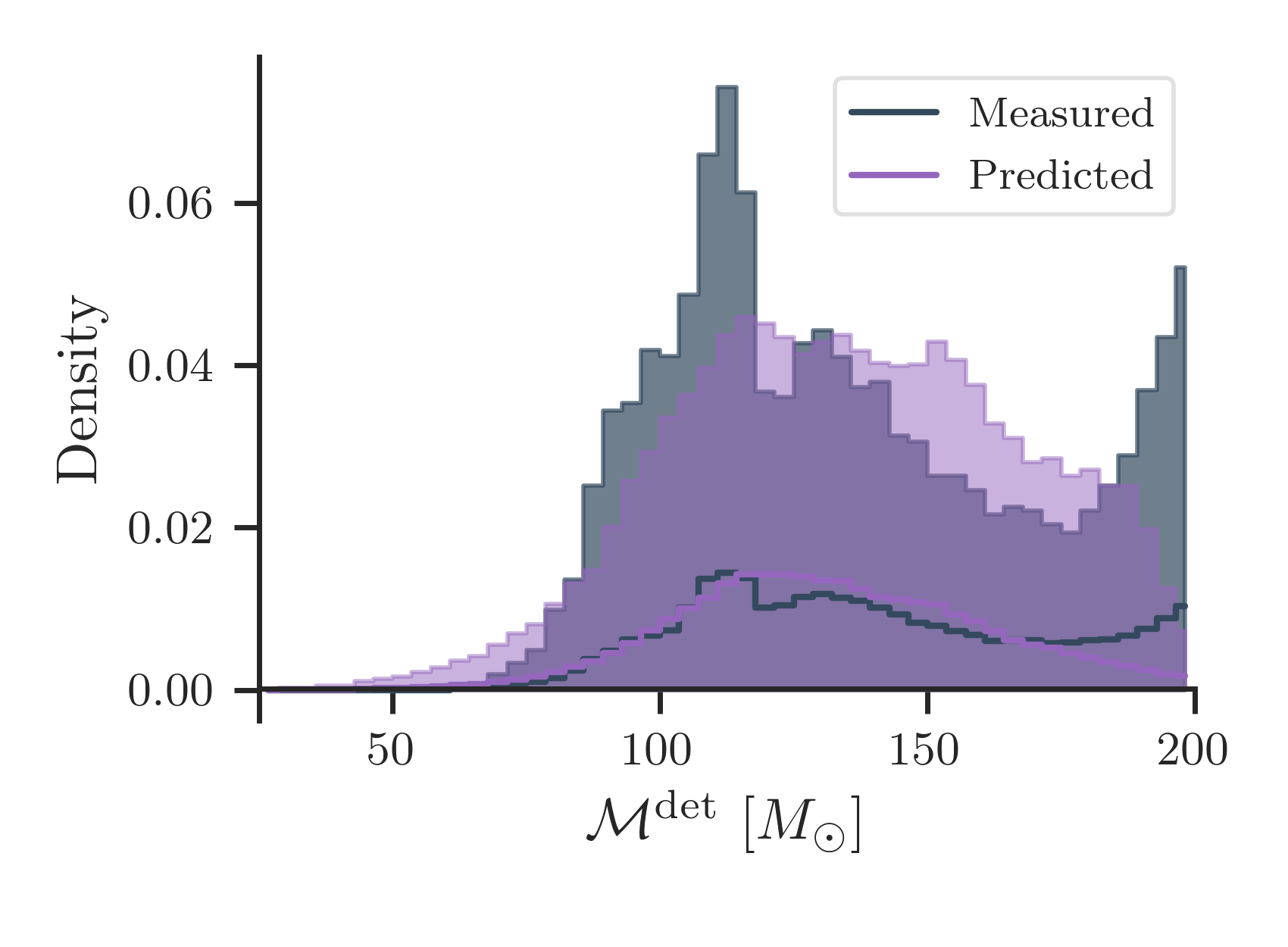}
         \\\hline
         H1 & Fast Scattering & \massratio & M2: Skew-Normal & \begin{tabular}{c}
$\mu=\HanfordfastscatteringskewnormalmuZeromassratiomedian_{-\HanfordfastscatteringskewnormalmuZeromassratiominus}^{+\HanfordfastscatteringskewnormalmuZeromassratioplus}$\\
$\sigma=\HanfordfastscatteringskewnormalsigmaZeromassratiomedian_{-\HanfordfastscatteringskewnormalsigmaZeromassratiominus}^{+\HanfordfastscatteringskewnormalsigmaZeromassratioplus}$\\
$\kappa=\HanfordfastscatteringskewnormalkappaZeromassratiomedian_{-\HanfordfastscatteringskewnormalkappaZeromassratiominus}^{+\HanfordfastscatteringskewnormalkappaZeromassratioplus}$\\
         \end{tabular} &        
         \ici{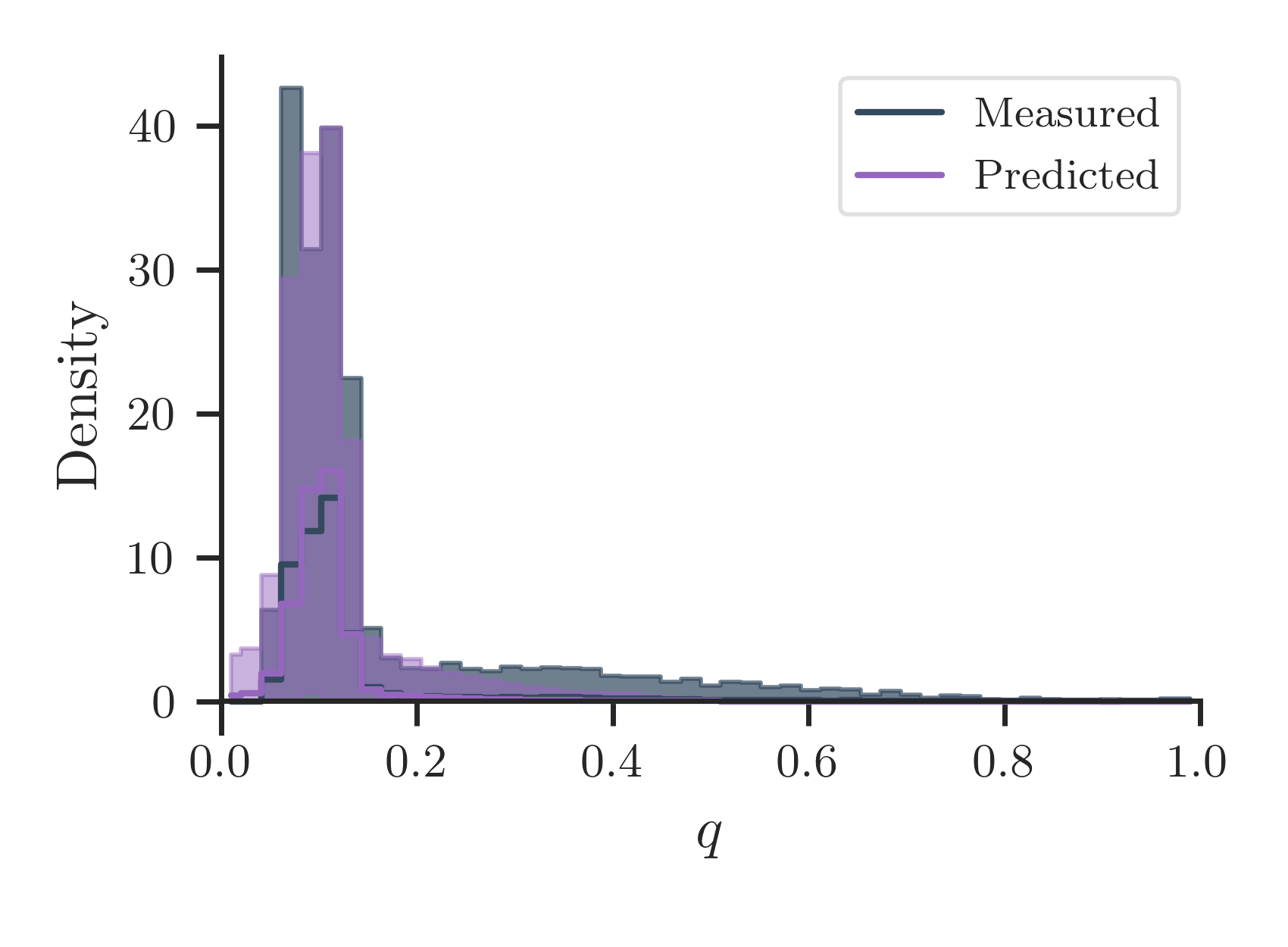}
         \\\hline
         L1 & Fast Scattering & \massratio & M2: Skew-Normal & \begin{tabular}{c}
$\mu=\LivingstonfastscatteringskewnormalmuZeromassratiomedian_{-\LivingstonfastscatteringskewnormalmuZeromassratiominus}^{+\LivingstonfastscatteringskewnormalmuZeromassratioplus}$\\
$\sigma=\LivingstonfastscatteringskewnormalsigmaZeromassratiomedian_{-\LivingstonfastscatteringskewnormalsigmaZeromassratiominus}^{+\LivingstonfastscatteringskewnormalsigmaZeromassratioplus}$\\
$\kappa=\LivingstonfastscatteringskewnormalkappaZeromassratiomedian_{-\LivingstonfastscatteringskewnormalkappaZeromassratiominus}^{+\LivingstonfastscatteringskewnormalkappaZeromassratioplus}$\\
         \end{tabular} &        
         \ici{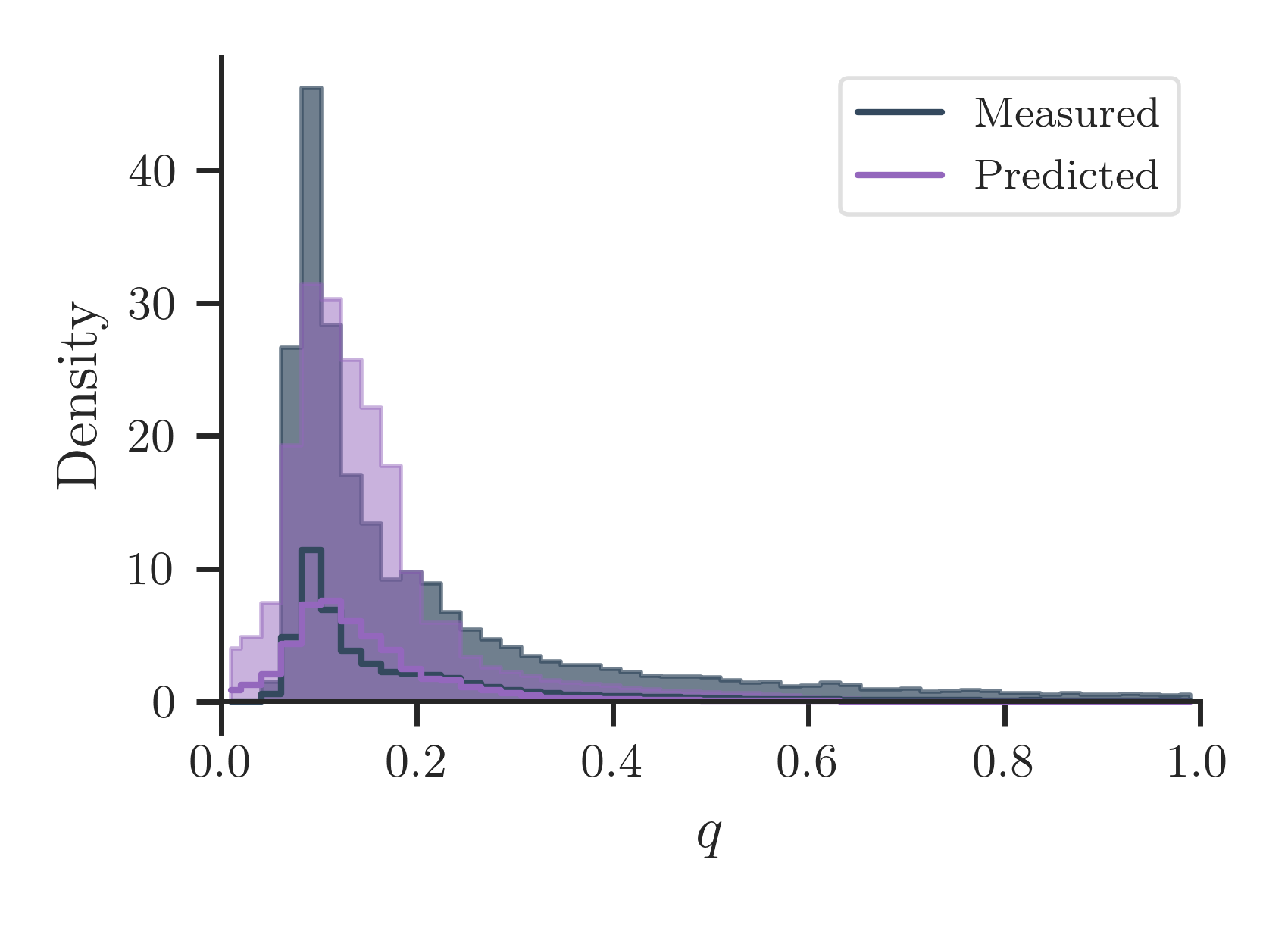}
         \\\hline
         H1 & Fast Scattering & \primaryspin & M3: Powerlaw & \begin{tabular}{c}
$\alpha=\HanfordfastscatteringpowerlawalphaZeroaOnemedian_{-\HanfordfastscatteringpowerlawalphaZeroaOneminus}^{+\HanfordfastscatteringpowerlawalphaZeroaOneplus}$\\
         \end{tabular} &        
         \ici{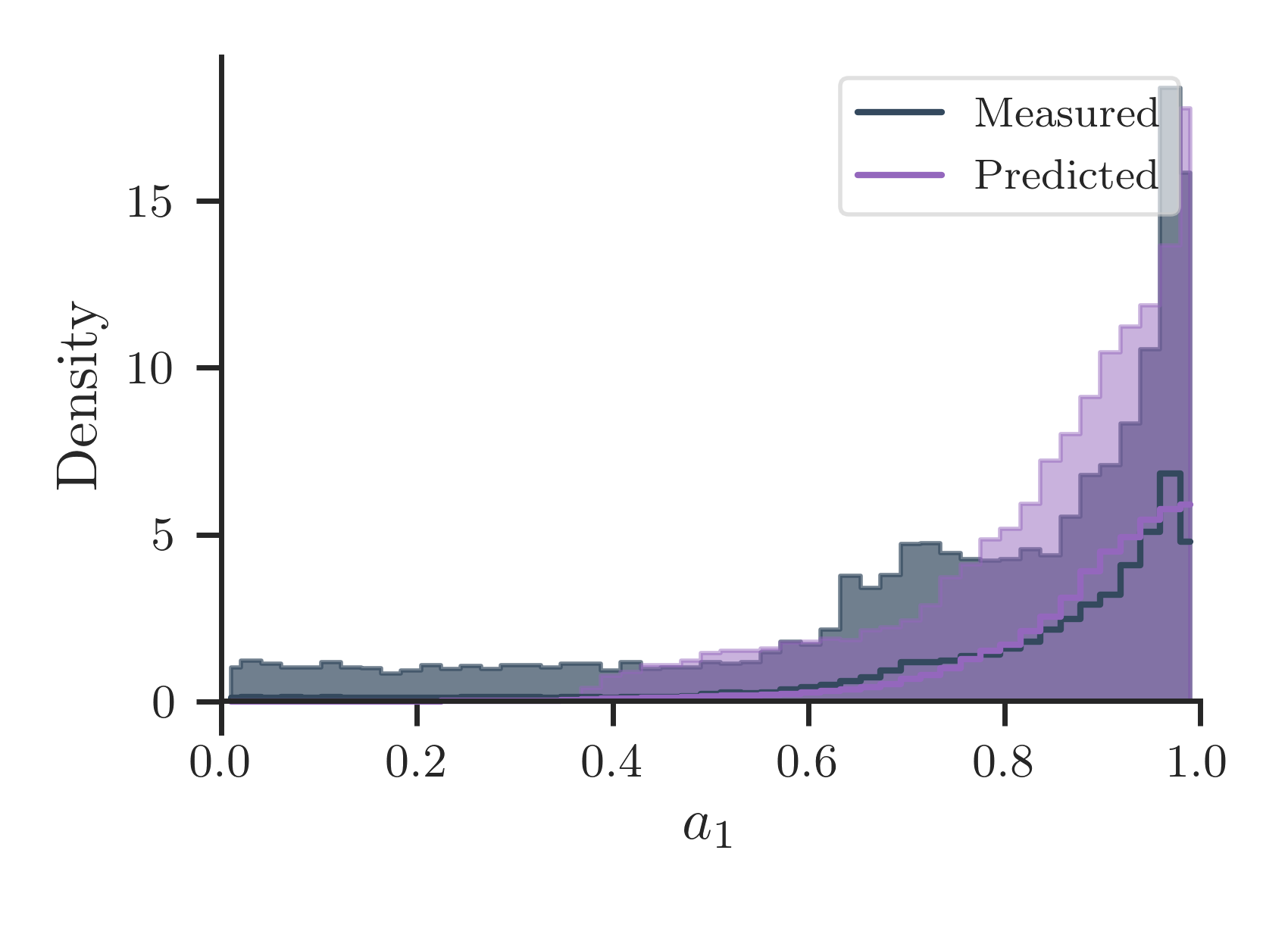}
         \\\hline
         L1 & Fast Scattering & \primaryspin & M3: Powerlaw & \begin{tabular}{c}
$\alpha=\LivingstonfastscatteringpowerlawalphaZeroaOnemedian_{-\LivingstonfastscatteringpowerlawalphaZeroaOneminus}^{+\LivingstonfastscatteringpowerlawalphaZeroaOneplus}$\\
         \end{tabular} &        
         \ici{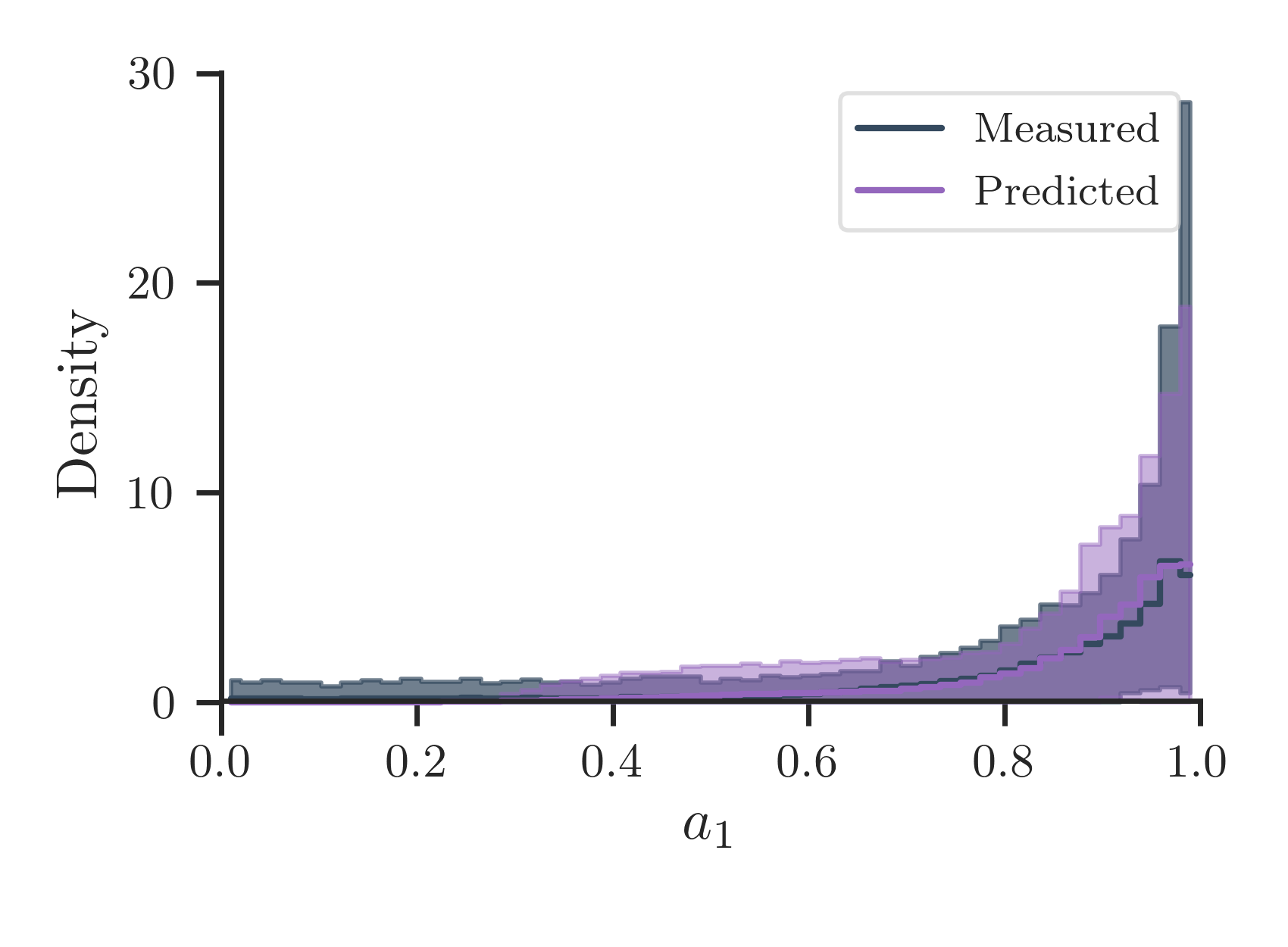}
         \\\hline
    \end{tabular}
    \caption{Table of parameterised models for Fast Scattering glitches in the H1 and L1 interferometers during the O3a era. Uncertainties on the median denote the symmetric 90\% credible interval. See Sec.~\ref{sec:methodology} for a description of the posterior predictive plots.}
    \label{tab:fast_scattering}
\end{table*}

\begin{table*}[h]
    \centering
    \begin{tabular}{c|c|c|c|l|c}
         Detector & Type & Parameter & Model & Median & Posterior Predictive  \\ \hline\hline
         H1 & Scattered Light & \mchirp & M2: Skew-Normal & \begin{tabular}{c}
$\mu=\HanfordscatteredlightskewnormalmuZerochirpmassmedian_{-\HanfordscatteredlightskewnormalmuZerochirpmassminus}^{+\HanfordscatteredlightskewnormalmuZerochirpmassplus}$\\
$\sigma=\HanfordscatteredlightskewnormalsigmaZerochirpmassmedian_{-\HanfordscatteredlightskewnormalsigmaZerochirpmassminus}^{+\HanfordscatteredlightskewnormalsigmaZerochirpmassplus}$\\
$\kappa=\HanfordscatteredlightskewnormalkappaZerochirpmassmedian_{-\HanfordscatteredlightskewnormalkappaZerochirpmassminus}^{+\HanfordscatteredlightskewnormalkappaZerochirpmassplus}$\\
         \end{tabular} & 
         \ici{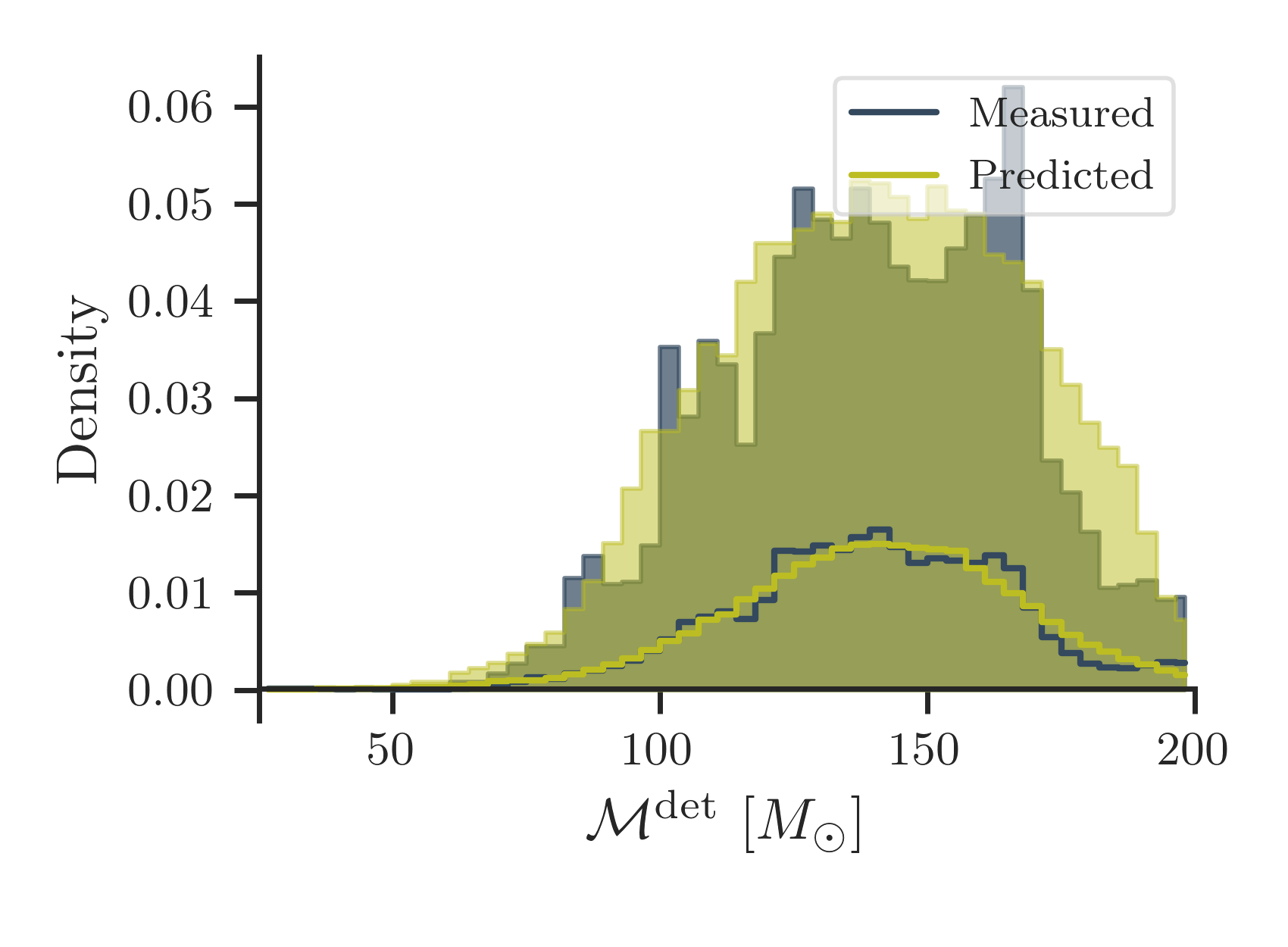}
         \\\hline
         L1 & Scattered Light & \mchirp & M2: Skew-Normal & \begin{tabular}{c}
$\mu=\LivingstonscatteredlightskewnormalmuZerochirpmassmedian_{-\LivingstonscatteredlightskewnormalmuZerochirpmassminus}^{+\LivingstonscatteredlightskewnormalmuZerochirpmassplus}$\\
$\sigma=\LivingstonscatteredlightskewnormalsigmaZerochirpmassmedian_{-\LivingstonscatteredlightskewnormalsigmaZerochirpmassminus}^{+\LivingstonscatteredlightskewnormalsigmaZerochirpmassplus}$\\
$\kappa=\LivingstonscatteredlightskewnormalkappaZerochirpmassmedian_{-\LivingstonscatteredlightskewnormalkappaZerochirpmassminus}^{+\LivingstonscatteredlightskewnormalkappaZerochirpmassplus}$\\
         \end{tabular} & 
         \ici{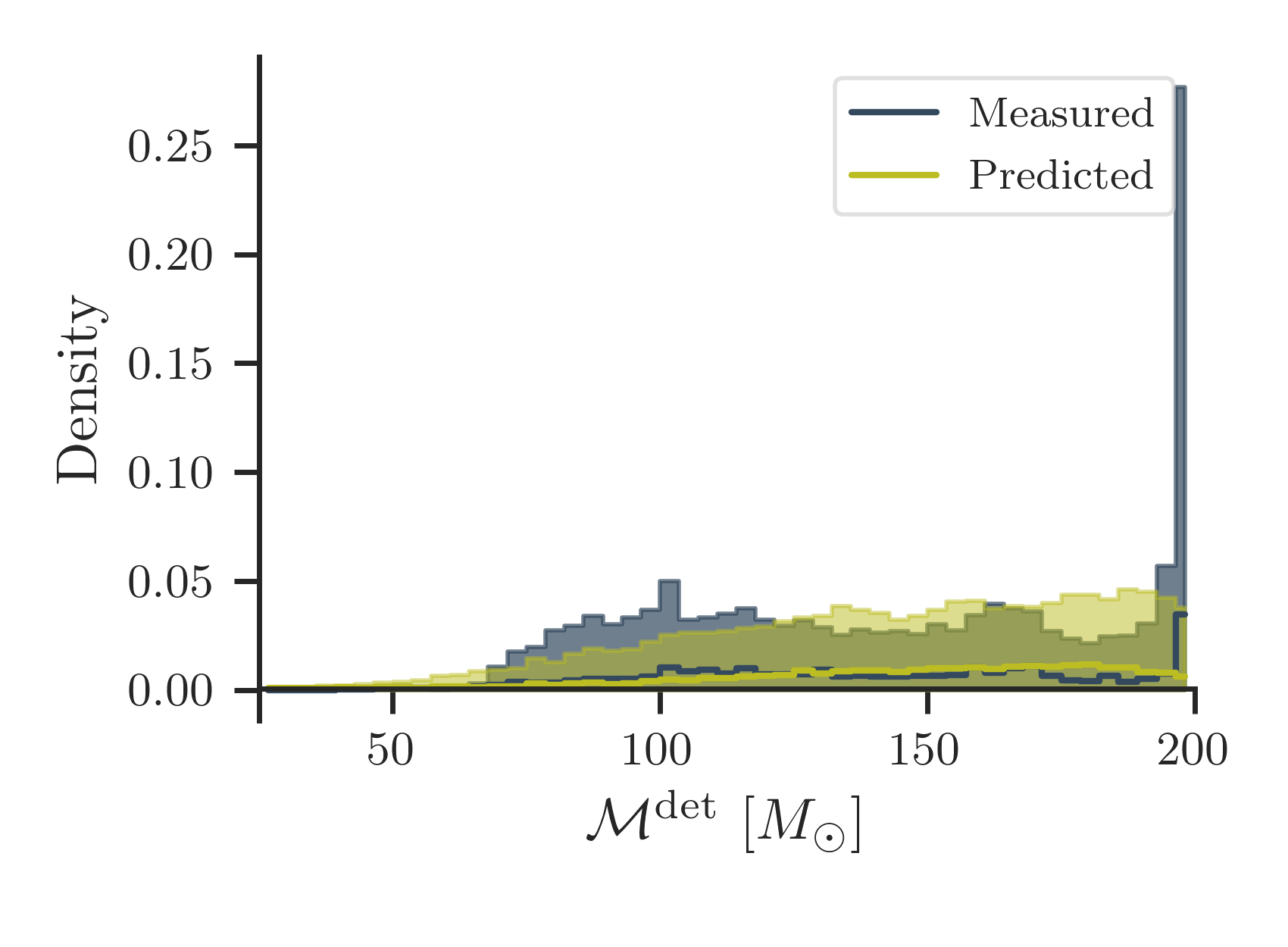}
         \\\hline
         H1 & Scattered Light & \massratio & M2: Skew-Normal & \begin{tabular}{c}
$\mu=\HanfordscatteredlightskewnormalmuZeromassratiomedian_{-\HanfordscatteredlightskewnormalmuZeromassratiominus}^{+\HanfordscatteredlightskewnormalmuZeromassratioplus}$\\
$\sigma=\HanfordscatteredlightskewnormalsigmaZeromassratiomedian_{-\HanfordscatteredlightskewnormalsigmaZeromassratiominus}^{+\HanfordscatteredlightskewnormalsigmaZeromassratioplus}$\\
$\kappa=\HanfordscatteredlightskewnormalkappaZeromassratiomedian_{-\HanfordscatteredlightskewnormalkappaZeromassratiominus}^{+\HanfordscatteredlightskewnormalkappaZeromassratioplus}$\\
         \end{tabular} &        
         \ici{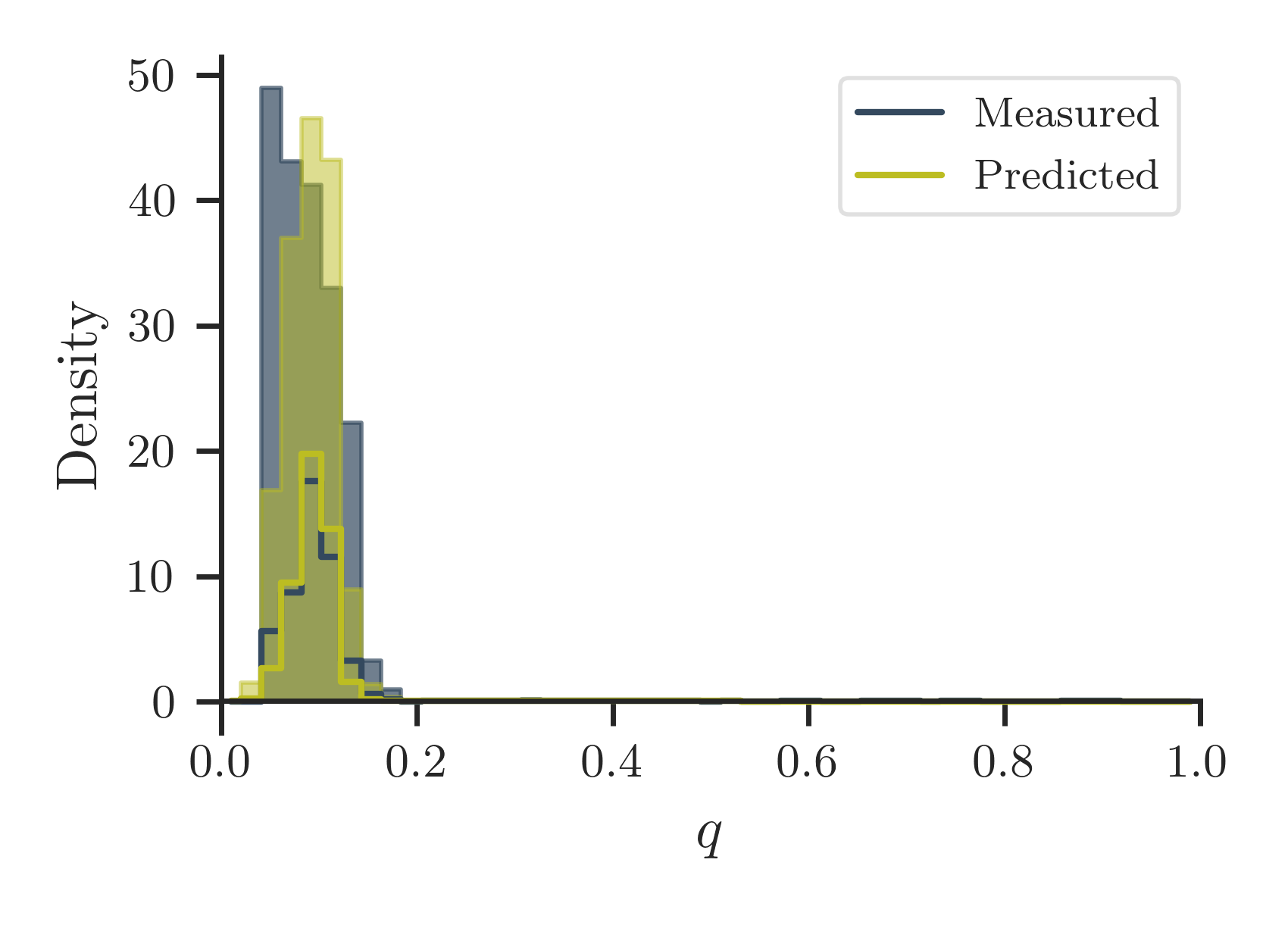}
         \\\hline
         L1 & Scattered Light & \massratio & M2: Skew-Normal & \begin{tabular}{c}
$\mu=\LivingstonscatteredlightskewnormalmuZeromassratiomedian_{-\LivingstonscatteredlightskewnormalmuZeromassratiominus}^{+\LivingstonscatteredlightskewnormalmuZeromassratioplus}$\\
$\sigma=\LivingstonscatteredlightskewnormalsigmaZeromassratiomedian_{-\LivingstonscatteredlightskewnormalsigmaZeromassratiominus}^{+\LivingstonscatteredlightskewnormalsigmaZeromassratioplus}$\\
$\kappa=\LivingstonscatteredlightskewnormalkappaZeromassratiomedian_{-\LivingstonscatteredlightskewnormalkappaZeromassratiominus}^{+\LivingstonscatteredlightskewnormalkappaZeromassratioplus}$\\
         \end{tabular} &        
         \ici{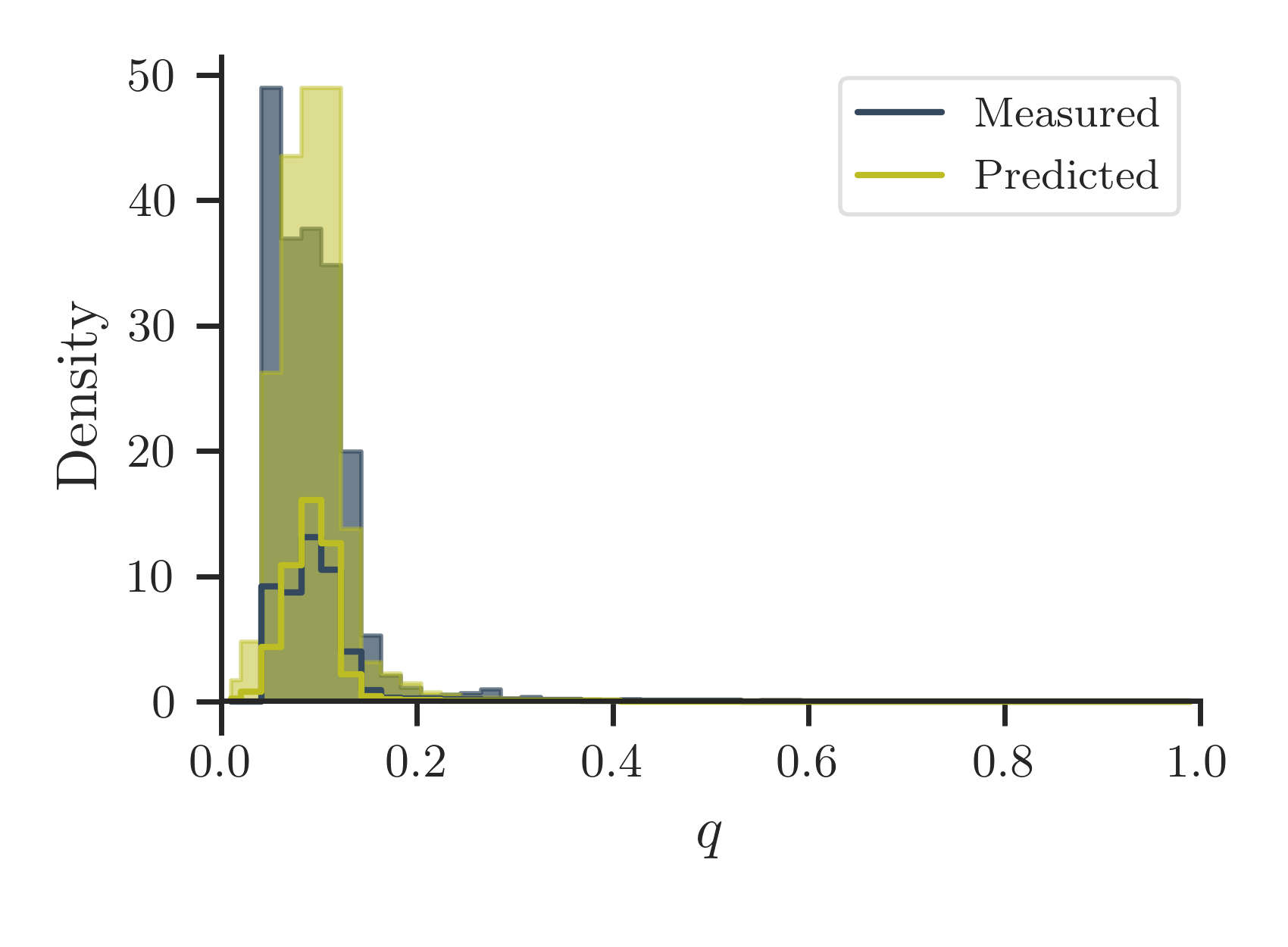}
         \\\hline
         H1 & Scattered Light & \primaryspin & M1: Powerlaw & \begin{tabular}{c}
$\alpha=\HanfordscatteredlightpowerlawalphaZeroaOnemedian_{-\HanfordscatteredlightpowerlawalphaZeroaOneminus}^{+\HanfordscatteredlightpowerlawalphaZeroaOneplus}$\\
         \end{tabular} &        
         \ici{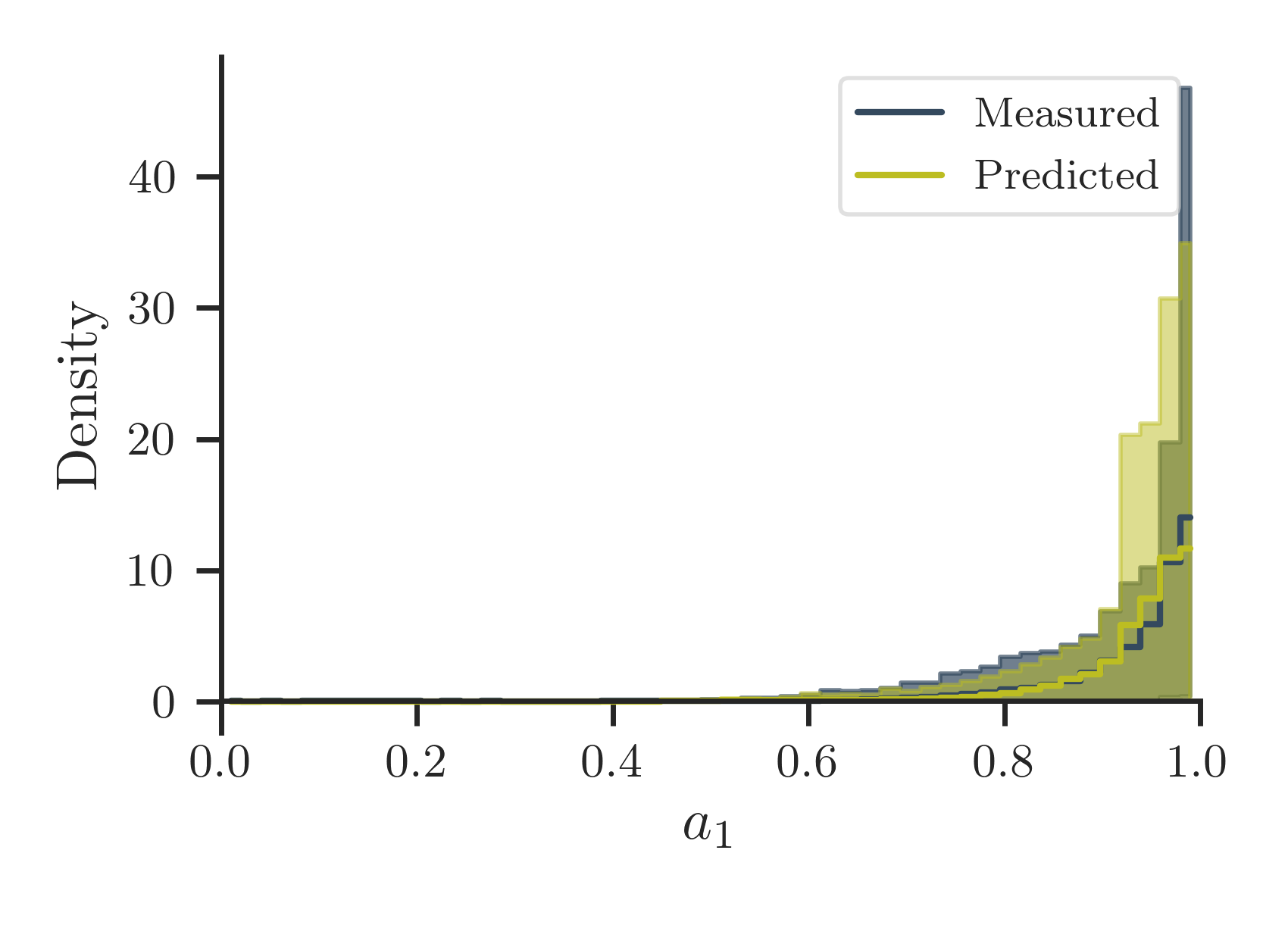}
         \\\hline
         L1 & Scattered Light & \primaryspin & M1: Powerlaw & \begin{tabular}{c}
$\alpha=\LivingstonscatteredlightpowerlawalphaZeroaOnemedian_{-\LivingstonscatteredlightpowerlawalphaZeroaOneminus}^{+\LivingstonscatteredlightpowerlawalphaZeroaOneplus}$\\
         \end{tabular} &        
         \ici{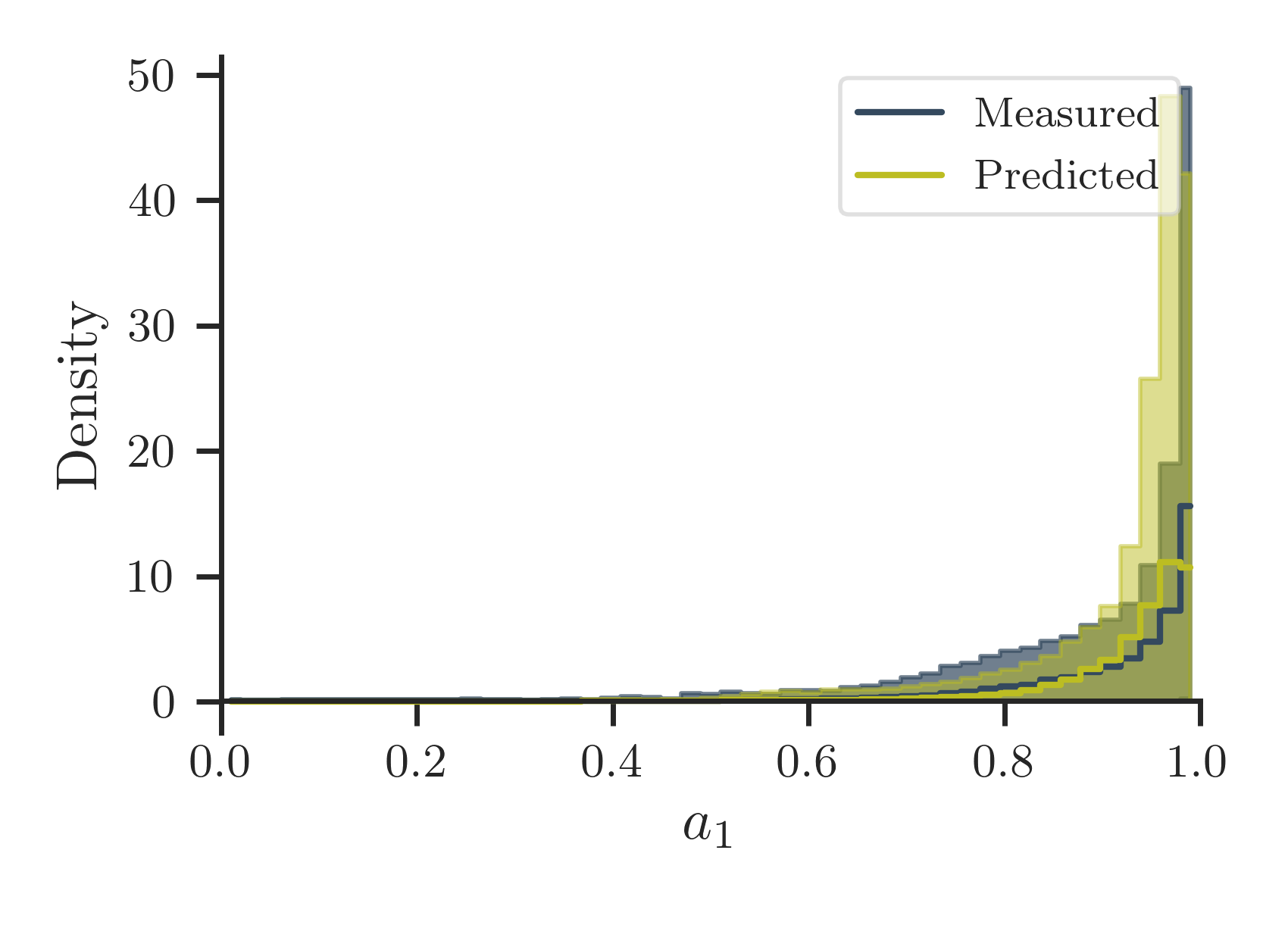}
         \\\hline
    \end{tabular}
    \caption{Table of parameterised models for scattered light glitches in the H1 and L1 interferometers during the O3a era. Uncertainties on the median denote the symmetric 90\% credible interval.}
    \label{tab:scattered_light}
\end{table*}
\end{document}